\documentclass[11pt,a4]{book}

\usepackage{amsmath}
\usepackage{graphicx}
\usepackage[superscript]{cite}

\newcommand{\dir}{Figs}

\newcommand{\kBT}{k_{_B}\!T}
\newcommand{\gD}{g_{_D}\!}

\begin{document}

\setcounter{chapter}{2}

\chapter[Theory and Simulation]{Theory and Simulation of Multiphase Polymer Systems}

\begin{center}
Friederike Schmid \\
{\em Institute of Physics, Johannes-Gutenberg Universit\"{a}t Mainz,
Germany}
\end{center}

\section{Introduction}

The theory of multiphase polymer systems has a venerable
tradition. The 'classical' theory of polymer demixing, the
Flory-Huggins theory, was developed already in the forties of the
last century \cite{flory_41, huggins_41}. It is still the starting
point for most current approaches -- be they improved theories for
polymer (im)miscibility that take into account the microscopic
structure of blends more accurately, or sophisticated field theories
that allow to study inhomogeneous multicomponent systems of polymers
with arbitrary architectures in arbitrary geometries. 
In contrast, simulations of multiphase polymer systems
are relatively young. They are still limited by the fact that one must
simulate a large number of large molecules in order to obtain
meaningful results. Both powerful computers and smart modeling and
simulation approaches are necessary to overcome this problem.

In the limited space of this chapter, I can only give a taste of
the state-of-the-art in both areas, theory and simulation. Since the
theory has reached a fairly mature stage by now, many aspects of it 
are covered in textbooks on polymer physics \cite{flory_book, 
degennes_book, doi_edwards_book, fleer_cohenstuart_book, larson_book,
fredrickson_book, hamley_book, fredrickson_book}. The information on
the state-of-the art of simulations is much more scattered. This is
why I have put some effort into putting together a representative
list of references in this area -- which is of course still far 
from complete.

The chapter is organized as follows. In Section \ref{sec:basic}, I
briefly introduce some basic concepts of polymer theory. The purpose
of this part is to make the chapter accessible to readers who are
not very familiar with polymer physics; it can safely be skipped by
the others.
Section \ref{sec:theory} is devoted to the theory of multiphase
polymer systems. I recapitulate the Flory-Huggins theory and
introduce in particular the concept of the Flory interaction
parameter (the $\chi$ parameter), which is a central ingredient in
most theoretical descriptions of multicomponent polymer systems.
Then I focus on one of the most successful mean-field theories for
inhomogeneous (co)polymer blends, the self-consistent field theory.
I sketch the main idea, discuss various aspects of the
theory and finally derive popular analytical approximations for
weakly and strongly segregated blends (the random phase
approximation and the strong-segregation theory).

In Section \ref{sec:simulations}, I turn to discussing simulations
of multiphase polymer systems. A central concept in this research
area is 'multiscale modeling': Polymers cannot be treated at all
levels of detail simultaneously, hence coarse-grained models are
used in order to study different aspects of the materials in
different simulations. This allows one to push the simulation limits
to larger length and time scales. I describe some of the most
popular coarse-grained structural and dynamical models and
give an overview over the state-of-the-art of simulations of polymer
blends and copolymer melts.

\section{Basic Concepts of Polymer Theory}

\label{sec:basic}

For the sake of readers who are not familiar with polymer theory, I
begin with recapitulating very briefly some basic concepts. 

Polymers are macromolecules containing up to hundreds of thousands of 
atoms.  At first sight, one would not expect such molecules to be 
easily amenable to theoretical modeling; however, it turns out that the 
large size of the molecules and their highly repetitive structure 
in fact simplifies things considerably. Since polymer molecules interact 
with many others, details of local interactions average out and polymers 
can often be characterized by a few effective quantities, such as their
topology, the local stiffness along the backbone, the bulkiness, the
compatibility/incompatibility of the building blocks etc. Already
decades ago, pioneers like Flory \cite{flory_book}, Edwards
\cite{doi_edwards_book}, de Gennes\cite{degennes_book} have
established theoretical polymer science as a highly successful field 
of research, which brings together scientists from theoretical
chemistry, statistical physics, materials science, and even the
biosciences, has created a wealth of new beautiful theoretical
concepts, and has not lost any of its fascination for theorists up
to date.

\subsection{Fundamental Properties of Polymer Molecules}

The characterizing property of {\em poly}mers is their highly
modular structure. They are composed of a large number of small
building blocks ({\em mono}mers), which are often all alike, but may
also be combined to arbitrary sequences (in the case copolymers and
biopolymers). The monomers are arranged in chains, which are usually
flexible on the nanometer length scale, {\em i.e.}, they can form
kinks at little energetic expense, they curve around and may assume
a large number of conformations at room temperature. The properties
of such flexible polymers are largely determined by the entropy of
the chain conformations. For example, the number of available
conformations is reduced if molecules are stretched, which leads to
a purely entropic restoring spring force\cite{guth_mark_34} (rubber
elasticity). Exposed to stress, polymeric systems respond by
molecular rearrangements, which takes time and results in
time-dependent strain (viscoelasticity).

The fundamental processes that govern the behavior of polymeric
materials do not depend on the chemical details of the monomer
structure. For qualitative purposes, polymer molecules can be
characterized by a few properties such as
\begin{itemize}
\item The architecture of the molecules (linear chains, rings, stars, etc.)
\item Physical properties (local chain stiffness, chain size, monomer volume)
\item Physicochemical properties
  (monomer sequence, compatibility, charges)
\item Special properties
  ({\em e.g.}, a propensity to develop crystalline or liquid crystalline
   order).
\end{itemize}

\subsection{Coarse-Graining, Part I}

\label{sec:coarse-graining_1}

The notion of 'coarse-graining' has lately become a buzzword in
materials science, but the underlying concept is actually quite old
in polymer science. The need for coarse-graining results from the
fact that polymeric materials exhibit structure on very different
length scales, ranging from Angstrom (the monomeric scale) to
hundreds of nanometers (typical molecule extensions) or micrometers
(supramolecular aggregates). It is not possible to treat all of them
within one common theoretical framework. Therefore, different
theoretical descriptions have been developed that deal with
phenomena on different length and time scales. On the {\em
microscale}, chemical details are taken into account and the
polymers are treated at an atomistic level. This is the realm of
theoretical chemistry. On the {\em mesoscale}, simplified molecule
models come into play (string models, lattice models, bead-spring
models, see below), whose behavior can be understood with concepts
from statistical physics. Finally, on the {\em macroscale},
polymeric materials are described by continuous fields (composition,
strain, stress etc.) with certain mechanical properties, and their
behavior can be calculated with methods borrowed from the engineers.

\begin{figure}
\centerline{
\includegraphics[width=0.48\textwidth]{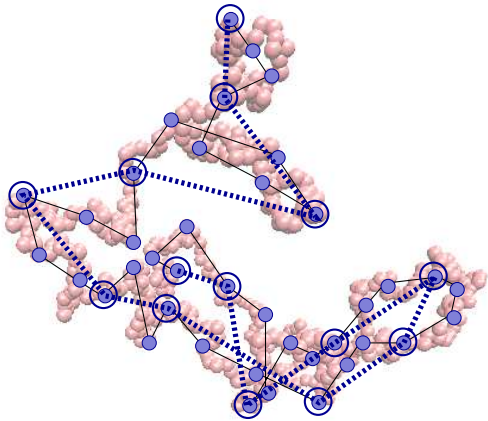}
} \caption{\label{fig:polymer-coarse} Mesoscopic coarse-grained
representations of a polymer molecule. Light pearl necklace in the
background: Bead chain with bonds of fixed length. Solid and dashed
lines, chain of coarse-grained units linked by bonds of variable
length. }
\end{figure}

In the following, I shall mainly focus on the mesoscale level,
where polymers are described by extended molecules made of
simplified ''monomeric'' units, each representing several real
monomers. Even within that level, one still has some freedom
regarding the choice of the coarse-grained units. This is
illustrated in Fig.\ \ref{fig:polymer-coarse}, where different
coarse-grained representations of a polymer are superimposed onto
each other. Polymers have remarkable universality properties, which
allow one to link different coarse-grained representations in a
rather well-defined manner, as long as the length scales under
consideration are much larger than the (chemical) monomer length
scale. For example, starting from one (atomistic or coarse-grained)
model, we can construct a coarse-grained model by combining $m$
''old'' units to one ''new'' unit. If $m$ is sufficiently large, the
average squared distance $\langle d^2 \rangle$ between two adjacent
new units will depend on $m$ according to a characteristic
power law
\begin{equation}
\label{eq:scaling}
\langle d^2 \rangle \sim m^{2\nu},
\end{equation}
where the exponent $\nu$ depends on the environment of a polymer,
but {\em not} on chemical details \cite{degennes_book,doi_edwards_book}.
In a dense polymer melt, one has $\nu = 1/2$ (see below). Similar
scaling laws can be established for other chain parameters.

\subsection{Ideal Chains}

In mesoscopic polymer theories, one often uses as a starting point a
virtual polymer chain where monomers that are well separated along
the polymer backbone do not interact with each other, even if their
spatial distance is small. Such polymers are called 'ideal chains'.
Even though they are mere theoretical constructions, they provide a
good approximative description of polymers in melts and in certain
solvents ('Theta'-solvents, see below).

\subsubsection{A Paradigm of Polymer Theory: The Gaussian Chain}

Let us now consider a flexible ideal chain with $N$ monomers, which
we coarse-grain several times as sketched in Fig.\
\ref{fig:polymer-coarse}, until one coarse-grained monomer unites
$m$ 'real' monomers. For large $m$, the resulting chain is a random
walk in space consisting of uncorrelated random steps ${\bf d}_i$ of
varying length. According to the central limit theorem of
probability theory \cite{paul_baschnagel_book}, the steps are
approximately Gaussian distributed,
${\cal P}({\bf d}) \sim \exp(- d^2 /2 m \sigma^2)$,
where $\sigma$ does not depend on $m$. The same random walk 
statistics can be reproduced by a Boltzmann distribution 
with an effective coarse-grained Hamiltonian
\begin{equation}\label{eq:hm}
{\cal H}_m = \frac{1}{2} \: \frac{\kBT}{m \sigma^2}
\sum_{i=1}^{N/m} {\bf d}_i^2.
\end{equation}
The Hamiltonian ${\cal H}_m$ describes the energy of a chain of
springs with spring constant $\kBT m/\sigma^2$. The coarse-graining
procedure has thus eliminated the information on chemical details
(they are now incorporated in the single parameter $\sigma$), and
instead unearthed the entropically induced elastic behavior of
the chain which lies at the heart of rubber elasticity. Eq.\
(\ref{eq:hm}) is also an example for universal behavior in a polymer
system (see section \ref{sec:coarse-graining_1}): The coarse-grained
chain is self-similar. Every choice of $m$ produces an equivalent
model, provided the spring constant is rescaled accordingly. The
distance between two coarse-grained units exhibits a scaling law of
the form (\ref{eq:scaling}) as a function of $m$, $\langle d^2
\rangle = 3 \sigma^2 m^{2\nu}$ with $\nu=1/2$.

Based on these considerations, it seems natural to define a
'generic' ideal chain model based on Eq.\ (\ref{eq:hm}) with $m=1$,
\begin{equation}\label{eq:hg_discrete}
\frac{{\cal H}_G[{\bf r}_i]}{\kBT} = \frac{1}{2 \sigma^2} \:
\sum_{i=1}^{N-1} ({\bf r}_{i+1}-{\bf r}_i)^2,
\end{equation}
the so-called 'discrete Gaussian chain' model. For theoretical purposes,
it is often convenient to take the continuum limit: The index $i$ in
Eq.\ (\ref{eq:hm}), which counts the monomers along the chain
backbone, is replaced by a continuous variable $s$, the chain is
parametrized by a continuous path ${\bf R}(s)$, and the steps ${\bf
d}$ correspond to the local derivatives ${\rm d}{\bf R}/{\rm d}s$ of
this path. The effective Hamiltonian then reads
\begin{equation}\label{eq:hg}
\frac{{\cal H}_G[{\bf R}]}{\kBT} = \frac{3}{2 b^2}
\int_0^N {\rm d}s \: \big(\frac{{\rm d}{\bf R}}{{\rm d}s}\big)^2.
\end{equation}
This defines the continuous Gaussian chain. The only material
parameters in Eq.\ (\ref{eq:hg})
are the chain length $N$ and the 'statistical
segment length' or 'Kuhn length' $b$. Even those two are
not independent, since they both depend on the definition of the
monomer unit. An equivalent chain model can be obtained by rescaling
$N \to N/\lambda$ and $b^2 \to b^2 \lambda$. Hence the only true
independent parameter is the extension of the chain, which can be
characterized by the squared gyration radius
\begin{equation}\label{eq:rg}
R_g^2 =
\frac{1}{N}
\int_{0}^N {\rm d}s \: ({\bf R}(s) - \overline{{\bf R}})^2
= b^2 N/6,
\end{equation}
where $\overline{{\bf R}} = 1/N \int {\rm d}s \: {\bf R}(s)$ is the
center of mass of the chain. The quantity $R_g$ sets the (only)
characteristic length scale of the Gaussian chain, and all length-dependent
quantities scale with $R_g$. For example, the structure factor 
is given by
\begin{equation}\label{eq:structure_factor}
S({\bf k}) = \frac{1}{N} \langle \left|
\int_0^N {\rm d}s \: e^{i {\bf k \: R}(s)} \right|^2 \rangle = N
\gD(k^2 R_g^2),
\end{equation}
with the Debye function
\begin{equation}\label{eq:debye}
\gD(x) = \frac{2}{x^2} (e^{-x} -  1 + x).
\end{equation}

The Gaussian chain is not only a prototype model for ideal chains,
it also provides a general framework for mesoscopic theories of
polymer systems. The Hamiltonian ${\cal H_G}$ (Eq.\ (\ref{eq:hg}))
is then supplemented by additional terms that account for
interactions, external fields, constraints ({\em e.g.}, chemical
crosslinks) etc. In this more general context, the Hamiltonian
(\ref{eq:hg}) is often referred to as 'Edwards Hamiltonian'.

Finally in this section, let us note that from a mathematical point of
view, the probability distribution of chain conformations defined by
Eq.\ (\ref{eq:hg}) is a Wiener measure \cite{paul_baschnagel_book,
kleinert_book}. The continuum limit leading to Eq.\ (\ref{eq:hg}) is
far from trivial, but well-defined. I shall not dwell further into
this matter.

\subsubsection{Other Chain Models}

The Gaussian chain model is a common starting point for analytical
theories of long flexible polymers on sufficiently large length
scales. On smaller length scales, or for stiffer polymers, or for
computer simulation purposes, other types of coarse-grained models
have proven useful. I briefly summarize some popular examples.
\begin{description}
\item The {\em wormlike chain} model is a continuous model designed
to describe stiff polymers. They are represented by smooth paths
${\bf R}(s)$ with fixed contour length $N$, where the parameter $s$
runs over the arc length of the curve, {\em i.e.}, the derivative
vector ${\bf u} = {\rm d}{\bf R}/{\rm d}s$ has length unity, $|{\bf
u}| \equiv 1$. The paths have a bending stiffness $\eta$, such that
they are distributed according to the effective Hamiltonian
\begin{equation}\label{eq:wlc}
\frac{{\cal H}_{WLC}[{\bf R}]}{\kBT} = \frac{\eta}{2} \int_0^N{\rm
d}s \: \big( \frac{{\rm d}^2{\bf R}}{{\rm d}s^2} \big)^2.
\end{equation}
The wormlike chain model is particularly useful if local orientational
degrees of freedom are important.
\item The {\em freely jointed} chain is a discrete chain model where
the chain is composed of $N$ links of fixed length. It is often used
to study general properties of ideal chains.
\item The {\em spring-bead} chain is a chain of beads connected with
springs. It has some resemblance with the discrete Gaussian chain
model, except that the springs have a finite equilibrium length.
Spring-bead models are popular in computer simulations.
\item In {\em lattice models}, the monomer positions are confined
to the sites of a lattice. This simplifies both theoretical
considerations and computer simulations.
\end{description}

\subsection{Interacting Chains}

The statistical properties of chains change fundamentally if
monomers interact with each other. Such interactions are readily
introduced in the coarse-grained models presented above. In the
discrete models, one simply adds explicit interactions between
monomers. In the continuous path models, one supplements the energy
contribution for individual ideal chains, Eq.\ (\ref{eq:hg}) or
(\ref{eq:wlc}), by an interaction term, such as
\begin{equation}\label{eq:hi}
{\cal H}_I[\hat{\rho}] =
\frac{v}{2} \int {\rm d}{\bf r} \: \hat{\rho}^2
+ \frac{w}{6} \int {\rm d}{\bf r} \: \hat{\rho}^3
+ \cdots
\end{equation}
(for weak interactions),
where the 'monomer density' $\hat{\rho}(\bf r)$ is defined as
\begin{equation}
\label{eq:rhohat}
\hat{\rho}({\bf r}) = \sum_\alpha \int_0^N {\rm d}s \:
\delta({\bf r} - {\bf R}_\alpha(s))
\end{equation}
and the sum $\alpha$ runs over the polymers ${\bf R}_\alpha(s)$
in the system.  Eq.\ (\ref{eq:hi}) corresponds to a virial expansion
of the local interaction energy in powers of the density. In many
cases, only the quadratic term ($v$) needs to be taken into account
('two-parameter Edwards model'). The Ansatz (\ref{eq:hi}) is
suitable for dilute polymer systems -- dense systems are discussed
below (Sec.~\ref{sec:melts}).  The generalization to multiphase
systems where monomers may have different type A,B,$\cdots$ is
straightforward. One simply operates with different densities
$\hat{\rho}_A, \hat{\rho}_B, \cdots$ and interaction parameters
$v_{AA}, v_{AB}, \cdots$. 

Interactions complicate the theoretical treatment considerably and
in general, exact analytical solutions are no longer available. The
properties of interacting polymer systems have been explored
theoretically within mean-field approximations,
renormalization-group calculations, scaling arguments, and computer
simulations. To set the stage for the discussion of multiphase
systems in sections \ref{sec:theory} and \ref{sec:simulations}, I
will now briefly sketch the most important scenarios for monophase
polymer systems.

\subsubsection{Polymers in Solution and Blobs}
\label{sec:solutions}

We first consider single, isolated polymer chains in solution. Their
properties depend on the quality of the solvent, which is
incorporated in the second virial parameter $v$ in Eq.~(\ref{eq:hi})
(the three-body parameter $w$ is typically positive\cite{osa_yoshizaki_04}).
In good solvent ($v>0$), monomers effectively repel each other, and the
chain swells. Extensive theoretical work \cite{degennes_book} has shown that
the scaling behavior (Eq.\ (\ref{eq:scaling})) remains valid, but
the exponent $\nu$ increases from $\nu=1/2$ for ideal chains to $\nu
\approx 3/(d+2)$, where $d$ is the spatial dimension (more
precisely, $\nu= 0.588$ in three dimensions). This is the famous
'Flory exponent', which characterizes the scaling behavior of
so-called 'self-avoiding chains'. Accordingly, the gyration radius
of the chain scales with the chain length like
\begin{equation}
\label{eq:rg_saw}
R_g^2 \sim N^{2 \nu}.
\end{equation}
In poor solvent ($v < 0$), monomers effectively attract each other
and the chain collapses. At the transition between the two regimes,
the 'Theta point' ($v_{\theta} \approx 0$), the scaling behavior
basically corresponds to that of ideal chains ($\nu = 1/2$), except
for subtle corrections due to the three-body $w$-term \cite{degennes_book}.

Eq.\ (\ref{eq:rg_saw}) describes
the behavior of single, unperturbed chains. Even in good
solvent, the self-avoiding scaling is often disturbed. For example,
the chains cannot swell freely if they are confined, or if they are
subject to external forces. Another important factor is the
concentration of chains in the solution: If many chains overlap, the
intrachain interactions are screened on large length scales. Loosely
speaking, monomers cannot distinguish between interactions with
monomers from the same chain and from other chains. As a result,
chains no longer swell and ideal chain behavior is recovered. This
mechanism applies in three or more spatial dimensions. Two
dimensional chains segregate \cite{degennes_book,
duplantier_saleur_87,semenov_03}.

\begin{figure}
\centerline{
\includegraphics[width=0.48\textwidth]{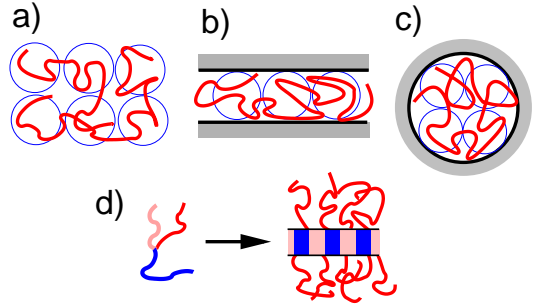}
}
\caption{\label{fig:blobs}
Illustration of the blob model in different situations:
a) concentrated polymer solution
b) Polymer confined in a slit
c) chain confined in a spherical cavity
d) structure formation in solutions of miktoarm star
copolymers (after Ref.\ \citeonline{zhulina_borisov_08}).
See text for explanation.
}
\end{figure}

All of these situations can be analyzed within one single ingenious
framework, the 'blob' picture introduced Daoud {\em et al.} in 1975
\cite{daoud_cotton_75}. It is based on the assumption that there 
exists a crossover length scale $\xi$ below which the chain is 
unperturbed. Blobs are
volume elements of size $\xi$ within which the polymers behave like
self-avoiding chains. On larger scales, the polymer behaves like an
ideal chain consisting of a string of blobs. Every blob contains $m
\sim \xi^{1/\nu}$ monomers and carries a free energy of the order
$\kBT$. These simple rules are the whole essence of the blob model.
I shall illustrate their use by applying them to a number of
prototype situations depicted in Fig.\ \ref{fig:blobs}.

\begin{description}
\item {\em Concentrated polymer solution} (Fig.\ \ref{fig:blobs} a).
For polymer concentrations $\phi$, we calculate the crossover length
scale $\xi$ from self-avoiding to ideal behavior. Since $\xi$ is the
blob size, we can simply equate $\phi = m/\xi^3$, {\em i.e.}, $\xi
\sim \phi^{-\nu/(3\nu-1)}$.
\item {\em Polymer confined in a slit} (Fig.\ \ref{fig:blobs} b).
We consider the free energy penalty $F$ on the confinement.
Here, the blob size is set by the width $R$ of the slit.
Each blob contains $m \sim R^{1/\nu}$ monomers, hence the total
free energy scales like $F \sim N/m \sim N R^{-1/\nu}$.
\item {\em Polymer confined in a cavity} (Fig.\ \ref{fig:blobs} c)
The result b) also holds for chains confined in a tube.
In closed cavities, however, the situation is different due
to the fact that the cavity constrains the monomer concentration.
The resulting blob size is $\xi \sim (N/R^3)^{\nu/(1-3 \nu)}$,
and the free energy of confinement scales as
$F \sim (R/\xi)^3 \sim (R/N^\nu)^{3/(1-3 \nu)}$.
This has been discussed controversially, but was recently confirmed
by careful computer simulations \cite{cacciuto_luijten_06}.
\item {\em ABC miktoarm star copolymers in selective solvent}
(Fig.\ \ref{fig:blobs} d).  Last, I cite a recent application to a
multiphase polymer system. Zhulina and Borisov\cite{zhulina_borisov_08}
have studied ABC star copolymers by means of scaling arguments. They
derived a rich state diagram, according to which ABC star copolymers
may assemble to several types of nanostructures, among other
spherical micelles, dumbbell micelles, and striped cylindrical
micelles. This is only one of numerous examples where scaling
arguments have been used to analyze complex multicomponent systems.
\end{description}

\subsubsection{Dense Melts}

\label{sec:melts}

Dense melts can be considered as extreme cases of a very
concentrated polymer solution, hence it is not surprising that the
chains effectively exhibit ideal chain behavior. In fact, the
situation is more complicated than this simple argument suggests.
The quasi-ideal behavior results from a cancellation of two effects:
On the one hand, the intrachain interactions promote chain swelling,
but on the other hand, the chain pushes other chains aside
('correlation hole'), which in turn exert pressure and squeeze it.
Deviations from true ideal behavior can be observed, {\em e.g.}, at
the level of chain orientational correlations \cite{wittmer_meyer_04,
wittmer_beckrich_07}.  Nevertheless, the ideality assumption is a
good working hypothesis in dense melts and shall also be used here
in the following.

\subsection{Chain Dynamics}

In this article I focus on equilibrium and static properties of
polymer systems. I can only touch on the possible dynamical
behavior, which is even more diverse.

In the time scales of interest, the motion of polymers is diffusive,
{\em i.e.}, the inertia of the macromolecules is not important.
Three prominent types of dynamical behavior have been established.
\begin{description}
\item In the {\em Rouse} regime, the chain dynamics is mainly
driven by direct intrachain interactions. This regime is encountered
for short chains. The dynamical properties of ideal chains can be
calculated exactly, and the results can be generalized to
self-avoiding chains using scaling arguments. One of the important
properties of Rouse chains is that their sedimentation mobility does
not depend on the chain length $N$. Hence the diffusion constant
scales like $D \sim 1/N$, and the longest internal relaxation time,
which can be estimated as the time in which the chain diffuses a
distance $R_g$, scales like $\tau \sim N^{2 \nu + 1}$.
\item In the {\em Zimm} regime, the dynamics is governed
by long-range hydrodynamic interactions between monomers. This
regime develops for sufficiently long chains in dilute solution.
They diffuse like Stokes spheres with the diffusion constant $D \sim
1/R_g$, and the longest relaxation time scales like $\tau \sim
R_g^3$. In concentrated solutions, the hydrodynamic interactions are
screened \cite{doi_edwards_book} and Rouse behavior is recovered after
an initial Zimm period \cite{ahlrichs_everaers_01}.
\item The {\em reptation} regime is encountered in dense systems
of chains with very high molecular weight. In this case, the chain
motion is topologically constrained by the surrounding polymer
network, and they are effectively confined to move along a tube in a
a snake-like fashion \cite{degennes_book,mcleish_02}. The diffusion
constant of linear chains scales like $D \sim 1/N^2$ and the longest
relaxation time like $\tau \sim N^3$.
\end{description}
This description is very schematic and oversimplifies the situation
even for fluids of linear polymers. Moreover, most polymer materials
are not in a pure fluid state. They are often cooled down below the
glass transition, or they partly crystallize -- in both cases, the
dynamics is frozen. Chemical or physical crosslinks constrain the
motion of the chains and impart solid-like behavior. In multiphase
polymer systems, the situation is further complicated by the fact
that the glass point or the crystallization temperature of the
different components may differ, such that one component freezes
where the other still remains fluid\cite{hu_mathot_03,
hu_05,hu_frenkel_05, wang_hu_06,ma_hu_07,ma_zha_08,mehta_kyu_04,
zhou_shi_08, khalatur_khokhlov_94,genix_arbe_05}.
The following discussion shall be limited to fluid multiphase 
polymer systems.

\section{Theory of Multiphase Polymer Mixtures}

\label{sec:theory}

After this general overview, I turn to the discussion of polymer
blends. We consider dense mixtures, where the polymers are in the
melt regime (Sec.\ \ref{sec:melts}). Moreover, we assume
incompressibility -- the characteristic length scales of density
fluctuations are taken to be much smaller than the length scales of
interest here.

Monomers of different type are usually slightly incompatible (see
Section \ref{sec:flory_huggins_parameter}). In polymers, the
incompatibilities are amplified, such that macromolecules of
different type tend to be immiscible: Blended together, they demix
and develop an inhomogeneous multiphase structure where
microdroplets of one phase are finely dispersed in another phase.

\begin{figure}
\centerline{
\includegraphics[width=0.7\textwidth]{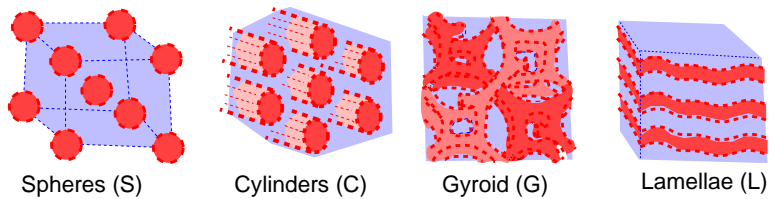}
}
\caption{\label{fig:structures_cops}
Self-assembled copolymer mesophases.
}
\end{figure}

In order to overcome or at least control this situation, copolymer
molecules can be added in which the two incompatible components are 
chemically linked to each other. They act as compatibilizer, 
{\em i.e.}, they shift the demixing transition and reduce the 
interfacial tension between different phases in the demixed region. 
At high concentrations, they are found to self-organize into a variety 
of ordered mesophases (microphase separation; see, {\em e.g.}, 
structures shown in Fig.\ \ref{fig:structures_cops}). Hence 
copolymers can also be used to manufacture nanostructured materials 
in a controlled way.

Nowadays, the theory of structure formation in polymer blends has
reached a highly advanced level and theoretical calculations have
predictive power, {\em e.g.}, with respect to structures that can be
expected in new polymeric materials. In this section, I shall
present some of the most successful theoretical approaches.

\subsection{Flory Huggins Theory}

I begin with sketching the Flory-Huggins theory, which is the
classical theory of phase separation in polymer blends, and which in 
some sense lays the foundations for all later, more sophisticated 
theories of polymer mixtures.

\subsubsection{Basic Model for Binary Blends}

\label{sec:flory_huggins}

We consider a binary blend of homopolymers A and B with length $N_A$
and $N_B$, and volume fractions $\Phi_A$ and $\Phi_B$. According to
Flory \cite{flory_41} and Huggins \cite{huggins_41}, the free energy
per monomer is approximately given by
\begin{equation}
\label{eq:flory_huggins}
\frac{f_{_{FH}}}{\kBT} =
\frac{\Phi_A}{N_A} \ln (\Phi_A)
+ \frac{\Phi_B}{N_B} \ln (\Phi_B)
+ \chi \Phi_A \Phi_B
\end{equation}
with $\Phi_A + \Phi_B = 1$. The first two terms account for the
mixing entropy of the two components, and the last term for the
(in)compatibility of the monomers. The parameter $\chi$ is the
famous 'Flory Huggins parameter', which will be discussed in more
detail below. The generalization of this expression to
ternary ABC homopolymer blends etc.\ is straighforward, one only
needs to introduce several $\chi$-parameters $\chi_{AB}$,
$\chi_{BC}$, and $\chi_{AC}$. Here we will only discuss binary
systems.

\begin{figure}
\centerline{
\includegraphics[width=0.8\textwidth,clip=true]{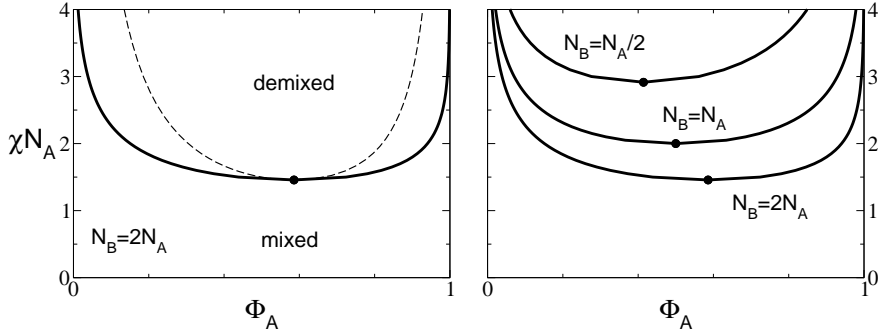}
} \caption{\label{fig:flory_huggins} 
Left: Phase diagram for a binary AB polymer blend with B-chains twice
as long as A-chains according to the Flory-Huggins theory. Thick solid
line shows the binodal line ({\em i.e.}, the demixing line), thin 
dashed line the spinodal line ({\em i.e.}, the line where the homogeneous
blend becomes unstable). 
Right: Binodals for binary AB blends with different
chain length ratios as indicated. }
\end{figure}

By minimizing the free energy, Eq.\ (\ref{eq:flory_huggins}), one
easily identifies the region in phase space where the mixture phase
separates into an A-rich phase and a B-rich phase.
At low values of $\chi$, the blend remains
homogeneous. Demixing sets in at a critical value $2\chi_c =
(1/\sqrt{N_A} + 1/\sqrt{N_B})^2$ for the critical composition
$\Phi_{A,c} = 1/(1 + \sqrt{N_A/N_B})$. The region of stability
of the homogeneous (mixed) blend is delimited by the 
''binodal'' line (see Fig.~\ref{fig:flory_huggins}).
Beyond the binodal, the homogeneous blend may still remain 
metastable.  It becomes unstable at the ''spinodal'', which is 
defined as the line where the second derivative of $f_{_{FH}}$ in 
Eq.~ (\ref{eq:flory_huggins}) with respect to $\Phi_A$ vanishes.
An example of a phase diagram with a binodal and a spinodal
is shown in Fig.~\ref{fig:flory_huggins} (left). 
Fig.~\ref{fig:flory_huggins} (right) demonstrates the shift
of the binodal with varying chain length ratio $N_A/N_B$. 

The Flory-Huggins free energy, Eq.\ (\ref{eq:flory_huggins}), was
originally derived based on a lattice model, but it can also be
applied to off-lattice systems. It does, however, rely on three
critical assumptions:
\begin{itemize}
\item The polymer conformations are taken to be those of ideal
chains, independent of the composition (ideality assumption, cf.
Sec.\ \ref{sec:melts}).
\item The melt is taken to be incompressible, and
monomers A and B occupy equal volumes.
\item Local composition fluctuations are neglected
(mean-field assumption).
\end{itemize}
In reality, none of these assumptions is strictly valid. The polymer
conformations do depend on the composition, most notably for chains
of the minority component. The incompressibility assumption is
reasonable, but the volumes per monomer are not equal. As a
consequence, the $\chi$-parameter is not a fixed parameter (at fixed
temperature), but depends on the composition of the blend (see Sec.\
\ref{sec:flory_huggins_parameter}). Finally, the composition
fluctuations shift phase boundaries and may even fundamentally
change the phase behavior. (see \ Sec.\ \ref{sec:fluctuations}).

\subsubsection{Inhomogeneous Systems: Flory-Huggins-de Gennes Theory}
\label{sec:flory_huggins_degennes}

Eq.\ (\ref{eq:flory_huggins}), only describes homogeneous systems. 
The simplest approach to generalizing the Flory-Huggins theory
to inhomogeneous systems, {\em e.g.}, polymer blends containing
interfaces, consists in adding a penalty on composition variations
$(\nabla \Phi_A)^2 = (\nabla \Phi_B)^2$. The coefficient of the
square gradient term can be derived within a more advanced
mean-field treatment, the random phase approximation, which will be
described further below (Sec.\ \ref{sec:RPA}). One obtains the
Flory-Huggins-de Gennes free energy functional for polymer blends,
\begin{equation}
\label{eq:flory_huggins_degennes}
{\cal F}_{\mbox{\tiny FHdG}}[\Phi_A({\bf r})]
= \rho_0 \int {\rm d}{\bf r}
\left\{
f_{_{FH}}(\Phi_A({\bf r}))
+ \frac{\kBT}{36}
\big( \frac{b_A^2}{\Phi_A} + \frac{b_B^2}{\Phi_B} \big)
(\nabla \Phi_A)^2
\right\}
\end{equation}
(with $\Phi_B = 1-\Phi_A$), where $b_A$ and $b_B$ are the Kuhn
lengths of the homopolymers A and B, and $f_{_{FH}}(\Phi_A))$ is
given by Eq.\ (\ref{eq:flory_huggins}). The functional
(\ref{eq:flory_huggins_degennes}) can be applied if composition
variations are weak, and have characteristic length scales of the
order of the gyration radius of the chains ('weak segregation
regime', see Sec.\ \ref{sec:wsl}).

A very similar functional can be derived in the opposite case,
where A- and B- polymers are fully demixed and separated by narrow
interfaces.  In this 'strong segregation' regime, the blend can be
described by the functional (see Sec.\ \ref{sec:ssl})
\begin{equation}
\label{eq:flory_huggins_ssl}
{\cal F}_{\mbox{\tiny SSL}}[\Phi_A({\bf r})]
= \rho_0 \int {\rm d}{\bf r}
\left\{ \chi \Phi_A \Phi_B + \frac{\kBT}{24}
\big( \frac{b_A^2}{\Phi_A} + \frac{b_B^2}{\Phi_B} \big)
(\nabla \Phi_A)^2
\right\}.
\end{equation}
At strong segregation, the mixing entropy terms 
in $f_{\mbox{\tiny FH}}$,  Eq.\ (\ref{eq:flory_huggins}), can be
neglected, hence the functionals (\ref{eq:flory_huggins_degennes})
and (\ref{eq:flory_huggins_ssl}) are identical except for the
numerical prefactor of the square gradient term. In the strong
segregation limit, the square gradient penalty results from an
entropic penalty on A and B {\em segments} due to the presence of
the interface, whereas in the weak segregation limit, it is
caused by the deformation of whole {\em chains}.

\subsubsection{Connection to Reality: The Flory-Huggins Parameter}
\label{sec:flory_huggins_parameter}

In the Flory-Huggins theory, the microscopic features of the blend
are incorporated in the single Flory-Huggins parameter $\chi$. Not
surprisingly, this parameter is very hard to access from first
principles.

In the original Flory-Huggins lattice model, $\chi$ is derived from
the energetic interactions between monomers that are neighbors on
the lattice. The interaction energy between monomers $i$ and $j$ is
taken to be characterized by energy parameters $\epsilon_{ij}$.
The $\chi$-parameter is then given by
\begin{equation}
\label{eq:flory_huggins_parameter}
\chi = \frac{z-2}{2\kBT} (
2\epsilon_{AB} - \epsilon_{AA} - \epsilon_{BB})
\end{equation}
where $z$ is the coordination number of the lattice. It is reduced
by two ($z-2$) in order to account for the fact that interactions
between neighbor monomers on the chain are fixed and have no
influence on the demixing behavior.

In reality, the situation is not as simple. The miscibility patterns
in real blends tend to deviate dramatically from that predicted by
the Flory-Huggins model. Several blends exhibit a lower critical
point instead of an upper critical point, indicating that the
demixing is driven by entropy rather than enthalpy. The critical
temperature $T_c$ of demixing often does not scale linearly with the
chain length $N$, as one would expect from Eq.
(\ref{eq:flory_huggins_parameter}) (see Fig.~\ref{fig:flory_huggins}). 
In blends of polystyrene and
poly(vinyl methyl ether) (PS/PVME), for example, $T_c$ is nearly
independent of $N$, and the critical concentration $\Phi_c$ is
highly asymmetric even for roughly equal chain lengths, in apparent
contrast with Fig.\ \ref{fig:flory_huggins} ($N_B=N_A$)
\cite{han_bauer_88}.

Formally, this problem can be resolved by arguing that the
Flory-Huggins expression for $\chi$,
Eq.~(\ref{eq:flory_huggins_parameter}), is oversimplified and 
that $\chi \kBT$ is not a constant. Several factors contribute to 
the $\chi$-parameter, leading to a complex dependence on the 
temperature, the blend composition, and even the chain length.
\begin{description}
\item[Monomer incompatibility]: Monomers may be incompatible both
for enthalpic and entropic reasons. For example, consider two nonpolar
monomers $i$ and $j$. The van-der Waals attraction between them is
proportional to the product of their polarizabilities $\alpha$,
hence $\epsilon_{ij} \propto \alpha_i \alpha_j$ and $\chi \propto
(\alpha_A - \alpha_B)^2$. More generally, the enthalpic
incompatibility of monomers $i$ and $j$ can be estimated by $\chi_H
\propto (\delta_A - \delta_B)^2$, where the $\delta_i$ are the
Hildebrand solubility parameters of the components
\cite{hildebrand_scott_book,biros_zeman_71}. In addition, entropic
factors may contribute to the monomer incompatibility, which are, 
{\em e.g.}, related to shape or stiffness disparities
\cite{freed_dudowicz_95, foreman_freed_98, dudowicz_freed_02,
dudowicz_freed_02b,freed_dudowicz_05}.
Since the enthalpic and entropic contributions evolve differently
as a function of the temperature, the $\chi$-parameter will in
general exhibit a complicated temperature dependence.
\item[Equation-of-state effect]: In general, the volume per
polymer depends on the composition of the blend. Already the volume
per monomer is usually different for different monomer species; at
constant pressure, $\chi$ therefore varies roughly linearly with
$\Phi_A$ \cite{kumar_00}. In blends of monomers with very similar
monomer structure, {\em e.g.}, isotopic blends, the linear contribution
vanishes and a weak parabolic dependence remains, which can partly
\cite{taylor_debenedetti_96} (but not fully \cite{melenkewitz_crist_00})
be explained by an `excess volume of mixing'.
\item[Chain correlations]: Since the demixing is driven
by intermolecular contacts, intramolecular contacts only contribute
indirectly to the $\chi$-parameter. The estimate for $\chi$ can be
improved if one replaces the factor $z-2$ by an `effective
coordination number' $z_{\mbox{\tiny eff}}$ which is given by the
mean number of interchain contacts per monomer
\cite{mueller_binder_95}. Moreover, the ideality assumption (see
Sec.\ \ref{sec:melts}) is not strictly valid. Chains in the minority
phase ({\em e.g.}, A chains in a B-rich phase) tend to shrink in
order to reduce unfavorable contacts \cite{mueller_binder_96}. Since
the segregation is effectively driven by $\chi N$, $\chi$ slightly
depends on the chain length $N$ as a result. The situation becomes
even more complex if copolymers are involved, which assume dumbbell
shapes even in a disordered environment \cite{fried_binder_91b,
fried_binder_91}. This also may affect the effective $\chi$-parameter
\cite{maurer_bates_98}.
\item[Composition correlations]: The effective interactions
between monomers change if the local environment is
not random. We have already noted earlier that the composition
may fluctuate.  Large-scale fluctuations can be incorporated
in the Flory-Huggins framework in terms of a fluctuating field
theory (see Secs.\ \ref{sec:fluctuations} and \ref{sec:wsl}).
Fluctuations (correlations) on the monomeric scale renormalize
the $\chi$-parameter ({\em nonrandom mixing}). Moreover, the local
fluid structure may depend on the local composition $\Phi$
({\em nonrandom packing}).
\end{description}

In view of these complications, establishing an exhaustive theory of
the $\chi$-parameter remains a formidable task \cite{foreman_freed_98,
freed_dudowicz_98, dudowicz_freed_00, curro_schweizer_88, schweizer_93,
schweizer_curro_97, wang_02}. Even the reverse problem of designing
simplified particle-based polymer models with a well-defined
$\chi$-parameters turns out to be highly non-trivial \cite{akkermans_08}.
The very concept of a $\chi$-parameter has been challenged repeatedly,
{\em e.g.}, by Tambasco {\em et al.} \cite{tambasco_lipson_06}, who
analyzed experimental data for a series of blends and found that their
thermodynamic behavior can be related to a single `$g-1$-parameter',
which is independent of composition, temperature and pressure. They
suggest that this parameter may be more appropriate to characterize
blends than the $\chi$-parameter. However, it has to be used in
conjunction with an integral equation theory, the BGY lattice theory
by Lipson \cite{lipson_91, lipson_92,tambasco_lipson_04}, which is
much more involved than the Flory-Huggins theory especially when
applied to inhomogeneous systems.

Freed and coworkers \cite{freed_pesci_87,freed_dudowicz_95,
foreman_freed_98,freed_dudowicz_05} have proposed a generalized
Flory-Huggins theory, the 'lattice cluster theory', which provides a 
consistent microscopic theory for macroscopic thermodynamic behavior. 
In a certain limit (high pressure, high molecular weight, fully
flexible chain), this theory reproduces a Flory-Huggins type free
energy with an effective $\chi$-parameter \cite{freed_dudowicz_98}
\begin{equation}
\chi_{\mbox{\tiny eff}} = \frac{(r_A-r_B)^2}{z^2} + \chi_0 \big( 1 -
\frac{2(p_A \Phi_A + p_B \Phi_B)}{z (z-2)} \big),
\end{equation}
where $\chi_0$ is given by Eq.\ (\ref{eq:flory_huggins_parameter})
and the parameters $r_j$, $p_j$ depend on the structure of the
monomers $j$. Dudowicz {\em et al.} have recently pointed out that
this model can account for a wide range of experimentally observed
miscibility behavior \cite{dudowicz_freed_02, dudowicz_freed_02b}.

Another pragmatic and reasonably successful approach consists in
using $\chi$ as a heuristic parameter. Assuming that it is at least
independent of the chain length, it can be determined
experimentally, {\em e.g.}, from fitting small-angle scattering data
to theoretically predicted structure factors
\cite{shibayama_yang_85,balsara_fetters_92, londono_wignall_97}.
Alternatively, $\chi$ can be estimated from atomistic simulations
\cite{fan_olafson_92,akkermans_08}. The results from experiment and
simulation tend to compare favorably even for systems of complex 
polymers \cite{kipper_seifert_04}.

\subsection{Self-consistent Field Theory}

\label{sec:scf}

I have taken some care to discuss the Flory-Huggins theory because
it establishes a framework for more general theories of polymer
blends. In particular, it provides the concept of the Flory-Huggins
parameter $\chi$, which will be taken for granted from now on, and
taken to be independent of the composition, despite the question
marks raised in the previous section (Sec.\ 
\ref{sec:flory_huggins_parameter}). In this section, I present a
more sophisticated mean-field approach, the self-consistent field
(SCF) theory. It was first proposed by Helfand and
coworkers\cite{helfand_tagami_71,helfand_tagami_72,helfand_tagami_72b,
helfand_sapse_75, helfand_75} and has since evolved to
be one of the most powerful tools in polymer theory. 
Reviews on the SCF approach can be found, {\em e.g.}, in Refs.\
\cite{schmid_98,matsen_02,fredrickson_book}.

\subsubsection{How it works in principle}

\label{sec:scf_principle}

For simplicity, I will first present the SCF formalism for
binary blends, and discuss possible extensions later.
Our starting point is the Edwards Hamiltonian for
Gaussian chains, Eq. (\ref{eq:hg}), with a Flory-Huggins
interaction term,
\begin{equation}
\label{eq:hi_binary}
{\cal H}_I[\hat{\rho}_A,\hat{\rho}_B]/\kBT = \rho_0 \: \chi \:
\int {\rm d}{\bf r} \: \hat{\Phi}_A \hat{\Phi}_B
\end{equation}
where we have defined the 'monomer volume fractions' $\hat{\Phi}_j =
\hat{\rho}_j/\rho_0$ and incompressibility is requested, {\em i.e.},
$\hat{\Phi}_A + \hat{\Phi}_B \equiv 1$ everywhere. The quantity
$\hat{\rho}_j$ depends on the paths ${\bf R}_\alpha$ of chains of
type $j$ and has been defined in Eq.\ (\ref{eq:rhohat}).

We consider a mixture of $n_A$ homopolymers A of length $N_A$ and
$n_B$ homopolymers B of length $N_B$ in the volume $V$.
The canonical partition function is given by
\begin{equation}
\label{eq:partition_1}
{\cal Z} =
\frac{1}{n_A! n_B!}
\left[ \prod_\alpha \rho_0 \int {\cal D}{\bf R}_\alpha \:
e^{- {\cal H}_G[{\bf R}_\alpha]/kBT} \right]
 e^{- \rho_0 \chi
\int {\rm d}{\bf r} \: \hat{\Phi}_A \hat{\Phi}_B} \:
\delta(\hat{\Phi}_A + \hat{\Phi}_B - 1).
\end{equation}
The product $\alpha$ runs over all chains in the system,
$\rho_0 \!\! \int {\cal D}{\bf R}_\alpha$ denotes the path integral
over all paths ${\bf R}_\alpha(s)$, and we have introduced the
factor $\rho_0$ in order to make ${\cal Z}$ dimensionless.

The path integrals can be decoupled by inserting delta-functions
$
\int {\cal D}{\rho}_j \: \delta(\rho_j-\hat{\rho}_j)
$
(with $j=$A,B), and using the Fourier representations of
the delta-functions
$ \delta(\rho_j-\hat{\rho}_j)
= \int_{i \infty} {\cal D}W_j \:
e^{\frac{1}{N} W_j (\rho_j - \hat{\rho}_j)}
$
and
$
\delta({\rho}_A + {\rho}_B - \rho_0)
= \int_{i \infty} {\cal D}\xi \:
e^{\frac{1}{N} \: \xi ({\rho}_A + {\rho}_B-\rho_0)}.
$
Here $N$ is some reference chain length, and
the factor $1/N$ is introduced for convenience.
This allows one to rewrite the partition function in the
following form:
\begin{equation}
\label{eq:partition_2} {\cal Z} \propto \int_{i \infty} {\cal D}W_A
\:{\cal D}W_B \: {\cal D}\xi \int_{\infty} {\cal D} \rho_A \: {\cal
D} \rho_B \: e^{- {{\cal F}}/{\kBT}}
\end{equation}
with
\begin{eqnarray}\label{eq:h_scf}
\frac{{\cal F[W_A,W_B,\xi,\rho_A,\rho_B]}_{\mbox{\tiny }}}{\kBT} &=
& \frac{\rho_0}{N} \: \Big\{ \: \chi N \!\int {\rm d}{\bf r} \:
\Phi_A \Phi_B - \sum_{j = A,B} \int \!{\rm d}{\bf r} \: W_j \Phi_j
\\ && \quad
- \int \!{\rm d}{\bf r} \: \xi \: (\Phi_A + \Phi_B - 1) - \!
\sum_{j= A,B} V_j \: \frac{N}{N_j} \: \ln( \rho_0 \: \frac{{\cal
Q}_j}{n_j} ) \Big\}, \nonumber
\end{eqnarray}
($\Phi_j=\rho_j/\rho_0$), where $V_j$ denotes the partial volume
occupied by all polymers of type $j$ in the system, and
the functional
\begin{equation}
\label{eq:part_q}
{\cal Q}_j = \int {\cal D}{\bf R}
\: e^{- {\cal H}_G[{\bf R}]/\kBT}
\:\: e^{-\frac{1}{N} \int_0^{N_j} {\rm d}s \: W_j({\bf R}(s))}.
\end{equation}
is the partition function of a single, noninteracting chain $j$ in the
external field $W_j$. Thus the path integrals are decoupled as intended,
and the coupling is transferred to the integral over fluctuating
fields $W_j$ and $\xi$.

Now the self-consistent field approximation consists in
replacing the integral (\ref{eq:partition_2}) by its
saddle point, {\em i.e.}, minimizing the effective
Hamiltonian ${\cal H}$ with respect to the variables
$\rho_j({\bf r})$ and $W_j({\bf r})$.
The minimization procedure results in a set of equations,
\begin{equation}
\label{eq:scf_equations}
\begin{array}{rclcccl}
\langle \hat{\Phi}_j \rangle &=& \Phi_j
&\quad &\mbox{with}&\quad& j=\mbox{A,B} \\
W_j &=& \chi N \langle \hat{\Phi}_i \rangle - \xi
&\quad & \mbox{with} &\quad& i,j=\mbox{A,B}  \quad \mbox{and} \quad i \ne j,
\end{array}
\end{equation}
where $\langle \hat{\Phi}_j \rangle$ denotes the average of
$\hat{\Phi}_j$ in a system of noninteracting chains subject to the
external fields $W_j({\bf r})$. One should note that the latter are real,
according to Eq.\ (\ref{eq:scf_equations}), even though the original
integral (\ref{eq:partition_2}) is carried out over the imaginary
axis. Intuitively, the $W_j({\bf r})$  can be interpreted as the 
effective mean fields acting on monomers due to the interactions 
with the surrounding monomers. Together with the incompressibility
constraint, the equations (\ref{eq:scf_equations}) form a closed
cycle which can be solved self-consistently.

For future reference, we note that it is sometimes convenient to
carry out the saddle point integral only with respect to the
variables $W_j({\bf r})$. This defines a free energy functional
${\cal F}_{\mbox{\tiny SCF}}[\Phi_A]$, which has essentially the
same form as ${\cal F}$ (Eq.\ \ref{eq:h_scf}), except that the
variables $W_\alpha({\bf r})$ and $\xi({\bf r})$ are now real
Lagrange parameter fields that enforce $\langle \hat{\Phi}_A \rangle 
= \Phi_A$, $\langle \hat{\Phi}_B \rangle = \Phi_B$, and 
$\Phi_B = 1-\Phi_A$ and depend on $\Phi_A$. The same functional can
also be derived by standard density functional approaches, using as
the reference system a gas of non-interacting Gaussian chains
\cite{freed_95}.

In some cases, one would prefer to operate in the grand canonical
ensemble, {\em i.e.}, at variable polymer numbers $n_j$. The
resulting SCF theory is very similar. The last term in Eq.\
(\ref{eq:h_scf}) is replaced by $(- \sum_j z_j {\cal Q}_j)$, where
$z_j$ is proportional to the fugacity of the polymers $j$.

The formalism can easily be generalized to other inhomogeneous
polymer systems. The application of the theory to multicomponent
A/B/C/... homopolymer blends with a more general interaction
Hamiltonian ${\cal H}_I
[\hat{\rho}_A,\hat{\rho}_B,\hat{\rho}_C,\cdots]$ replacing Eq.\
(\ref{eq:hi_binary}) is straightforward. The self-consistent field
equations (\ref{eq:scf_equations}) are simply replaced by
\begin{equation}
\label{eq:scf_fields}
W_j = \frac{N}{\kBT} \:
\frac{\delta {\cal H}_I[\{\rho_i\}]}{\delta \rho_j} - \xi.
\end{equation}
If copolymers are involved, the single-chain partition function
${\cal Q}_c$ for the corresponding molecules must be adjusted
accordingly. For example, the single-chain partition function for an
A:B diblock copolymer of length $N_c$ with A-fraction $f$ reads
\begin{equation}
{\cal Q}_c = \int {\cal D}{\bf R}
\: e^{- {\cal H}_G[{\bf R}]/\kBT}
\:\:
e^{\frac{1}{N} \int_0^{f N_c}   {\rm d}s \: W_A({\bf R}(s))
  +\frac{1}{N} \int_{f N_c}^{N_c} {\rm d}s \: W_B({\bf R}(s)) }.
\end{equation}
The general SCF free energy functional for incompressible
multicomponent systems is given by
\begin{eqnarray}
 \nonumber \frac{{\cal F}_{\mbox{\tiny SCF}}[\{\rho_j\}]}{\kBT} & = &
 \frac{{\cal H}_I[\{ \rho_i\}]}{\kBT} 
   -  \sum_j \frac{1}{N}\int {\rm d}{\bf r} \: W_j \: \rho_j 
   - \frac{1}{N} {\rm d}{\bf r} \: \xi (\sum_j \rho_j - \rho_0)
  \\ &&
   -   \sum_{\alpha} n_\alpha \: \ln( \rho_0 \:
  \frac{{\cal Q}_\alpha}{n_\alpha} ) , \label{eq:f_scf}
\end{eqnarray}
where the sum $j$ runs over monomer species and the sum $\alpha$
over polymer types.

The SCF theory has been extended in various ways to treat more
complex systems, {\em e.g.}, compressible melts and solutions
\cite{hong_noolandi_81,schmid_96}, macromolecules with complex
architecture \cite{matsen_schick_94b}, semiflexible polymers
\cite{morse_fredrickson_94} with orientational interactions
\cite{duechs_sullivan_02,hidalgo_sullivan_07},
charged polymers\cite{wang_taniguchi_04},
polydisperse systems\cite{cooke_shi_06,lynd_hillmyer_08},
polymer systems subject to stresses
\cite{hall_lookman_06,maniadis_lookman_07},
systems of polymers undergoing reversible bonds
\cite{feng_lee_97,lee_elliott_07},
or polymer/colloid composites \cite{thompson_ginzburg_01,
reister_fredrickson_05,sides_kim_06}.

\subsubsection{How it works in practice}

In the previous section, we have derived the basic equations of the
SCF theory. Now I describe how to solve them in practice. The first
task is to evaluate the single-chain partition functions ${\cal
Q}_j$ and the corresponding density averages $\langle \hat{\rho}_j
\rangle$ for noninteracting chains in an external field.

We consider a single ideal chain  of length $N$ in an external field
$W({\bf r},s)$, which may not only vary in space ${\bf r}$, but also
depend on the monomer position $s$ in the chain in the case of
copolymers. It is convenient to introduce partial partition
functions
\begin{eqnarray}
q({\bf r},s) &=& \int \!\! {\cal D}{\bf R} \:
\: e^{- {\cal H}_G[{\bf R}]/\kBT}
\:\: e^{\frac{1}{N} \int_0^s {\rm d}s' \: W({\bf R}(s'),s')}
\: \delta({\bf R(s)}-{\bf r}),
\\
q^\dag({\bf r},s) &=& \int \!\! {\cal D}{\bf R} \:
\: e^{- {\cal H}_G[{\bf R}]/\kBT}
\:\: e^{\frac{1}{N} \int_0^s {\rm d}s' \: W({\bf R}(s'),N-s')}
\: \delta({\bf R(s)}-{\bf r}),
\end{eqnarray}
where the path integrals ${\cal D} {\bf R}$ are carried out over
paths of length $s$. According to the Feynman-Kac formula
\cite{zinn_justin_book},
the functions $q({\bf r},s)$ and $q^\dag({\bf r},s)$ satisfy a
diffusion equation
\begin{eqnarray}
\label{eq:q_diffusion}
\frac{\partial}{\partial s} \: q({\bf r},s) &=&
(\frac{b^2}{6} \Delta - \frac{W({\bf r},s)}{N}) \: q({\bf r},s)
\\
\label{eq:qdag_diffusion}
\frac{\partial}{\partial s}\: q^\dag({\bf r},s) &=&
(\frac{b^2}{6} \Delta - \frac{W({\bf r}, N-s)}{N}) \: q^\dag({\bf r},s)
\end{eqnarray}
with the initial condition $q({\bf r},0) = q^\dag({\bf r},0)=1$.
Numerical methods to solve diffusion equations are
available\cite{mueller_schmid_05}, hence $q$ and $q^\dag$ are
accessible quantities. The single-chain partition function can then
be calculated {\em via}
\begin{equation}
{\cal Q}
= \int {\rm d}{\bf r} \: q({\bf r},N)
= \int {\rm d}{\bf r} \: q^\dag({\bf r},N)
\end{equation}
and the distribution of the $s$th monomer in space is
$ q({\bf r},s) \: q^\dag({\bf r},N-s)/{\cal Q}$.

More specifically, to study binary homopolymer blends, one must
solve the diffusion equations for the partial partition functions
$q_j =q^\dag_j$ in an external field $W = W_j$ (with $j =$ A,B). The
averaged volume fractions of monomers $j$ are then given by
\begin{equation}
\langle \hat{\Phi}_j ({\bf r}) \rangle = \frac{1}{\rho_0} \:
\frac{n_j}{{\cal Q}_j} \int_0^{N_j} {\rm d}s \: q_j({\bf r},s) \:
q^\dag_j({\bf r},N-s).
\end{equation}
in the canonical ensemble, and
\begin{equation}
\langle \hat{\Phi}_j ({\bf r}) \rangle =
\frac{z_j}{N} \:
\int_0^{N_j} {\rm d}s \: q_j({\bf r},s) \:
q^\dag_j({\bf r},N-s)
\end{equation}
in the grand canonical ensemble. If AB diblock copolymers with
fraction $f$ of A-monomers are present, one must calculate the
partial partition functions $q_c$ and $q^\dag_c$ in the external
field $W({\bf r},s) = W_A$ for $s < f N_c$ and $W({\bf r})= W_B$ for
$s \ge f N_c$. The contribution of the copolymers to the volume
fraction $\langle \hat{\Phi}_A \rangle$ is $n_C/\rho_0
{\cal Q}_C \: f_A({\bf r})$ with $f_A({\bf r}) = \int_0^{f N_c} {\rm
d}s\: q_C({\bf r},s) q_C^\dag({\bf r},N-s)$ in the canonical
ensemble, and $z_c/N f_A({\bf r})$ in the grandcanonical
ensemble.

With this recipe at hand, one can calculate the different terms in
Eqs.\ (\ref{eq:scf_equations}). The next problem is to solve these
equations simultaneously, taking account of the incompressibility
constraint. This is usually done iteratively. I refer the reader to
Section 3.4. in Ref.\ \citeonline{mueller_schmid_05} for a discussion
of different iteration methods.

\subsubsection{Application: Diblock Copolymer Blends, Part I}

\label{sec:diblocks_1}

\begin{figure}
\centerline{
\includegraphics[width=0.45\textwidth,clip=true]{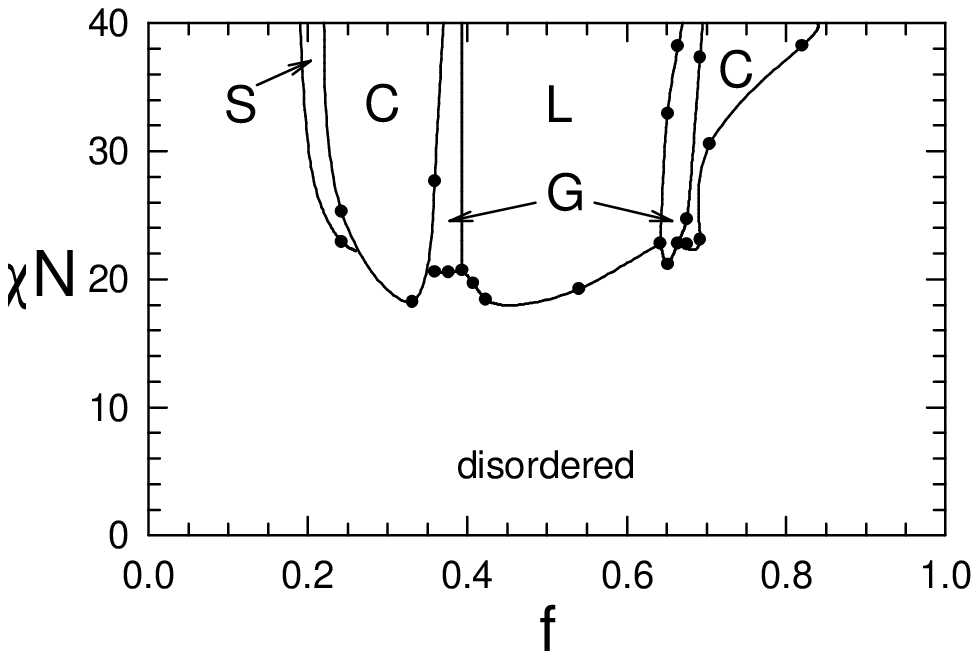}
\hspace{0.01\textwidth}
\includegraphics[width=0.45\textwidth,clip=true]{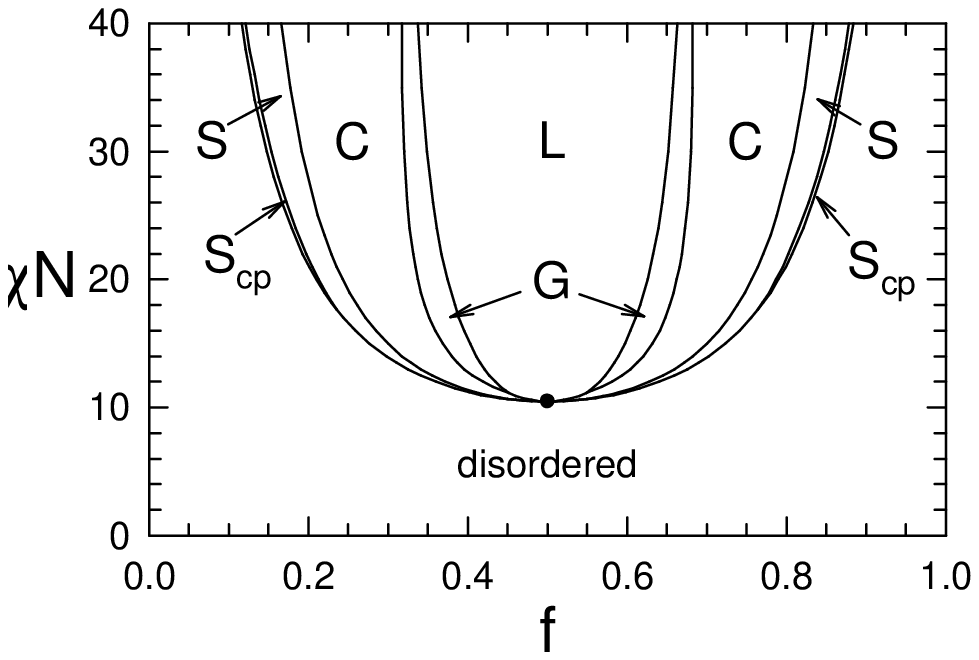}
}
\caption{\label{fig:diblock_1} Experimental phase diagram for
polystyrene-polyisoprene diblock copolymer melts (left, after Refs.\
\protect\citeonline{bates_schulz_94,foerster_khandpur_94,
khandpur_foerster_95}) compared with phase diagram obtained with
the SCF theory by Matsen and coworkers (right, after Ref.\
\protect\citeonline{matsen_bates_96}) in the coordinates $\chi N$ and
A block volume fraction $f$. The labels S,C,L, and G correspond to the
structures shown in Fig.\ \protect\ref{fig:structures_cops}. In addition,
the SCF phase diagram features a close-packed sphere phase S${}_{cp}$.
All phase transitions are first order except for the
disordered/lamellar-transition at $f=1/2$ in the SCF phase diagram.
Courtesy of Mark W. Matsen, adapted from M.W. Matsen, J. Phys.: Cond.
Matter {\bf 14}, R21 (2002).
}
\end{figure}

To illustrate the power of the SCF approach, I cite one of its most
spectacular successes: The reproduction of arbitrarily complex
copolymer mesophases.  In a series of seminal papers, Matsen and
coworkers have calculated phase diagrams for diblock copolymer
melts\cite{matsen_schick_94,matsen_bates_96}. Fig.\ \ref{fig:diblock_1} 
compares an experimental phase diagram due to Bates and coworkers
\cite{bates_schulz_94, foerster_khandpur_94, khandpur_foerster_95} 
with the SCF phase diagram of Matsen and Bates \cite{matsen_bates_96}. 
The SCF theory reproduces the experimentally observed structures. 
At high values of $\chi N$ (`strong segregation') the SCF phase diagram 
features the correct sequence of mesophases at almost the correct value 
of the fraction of A-monomers $f$. At low values of $\chi N$ ('weak
segregation'), the two phase diagrams are distinctly different. This
can be explained by the effect of fluctuations and will be
discussed further below (Secs.\ \ref{sec:fluctuations} and
\ref{sec:applications_copolymers}, see also Fig.\ \ref{fig:diblock_2}). 

Tyler and Morse have recently reconsidered the SCF phase diagram and 
predicted the existence of yet another mesophase, which has an orthorombic 
unit cell and an Fddd structure and intrudes in a narrow regime at the 
low $\chi N$-end of the gyroid phase \cite{tyler_morse_05}. This phase 
was later indeed found in a polystyrene-polyisoprene diblock copolymer 
melt by Takenaka and coworkers \cite{takenaka_wakada_07}.

\subsubsection{Related Mean-Field Approaches}
\label{eq:scf_other}

So far, I have focussed on sketching a variant of the SCF theory
which was originally developed by Helfand and coworkers \cite{helfand_75}.
A number of similar approaches have been proposed in the literature.

Scheutjens and Fleer have developed a SCF theory for lattice
models\cite{scheutjens_fleer_79}, which is applied very widely 
\cite{fleer_cohenstuart_book}. Scheutjens-Fleer calculations
are very efficient and incorporate in a natural way the finite
(nonzero) range of monomer interactions. To account for this in the
Helfand theory, one must introduce additional terms in (\ref{eq:hi})
\cite{helfand_sapse_75,helfand_75}, which indeed turn out to become
important in the vicinity of surfaces\cite{schmid_96}.

Carignano and Szleifer\cite{carignano_szleifer_93} have proposed a
SCF theory where chains are sampled as a whole in a surrounding mean
field. Hence intramolecular interactions are accounted for exactly and
the chain statistics corresponds to that of self-avoiding walks (Sec.\
\ref{sec:solutions}). This approach is more suitable than the
standard SCF theory to study polymers in solution, or melts of
molecules with low-molecular weight, where the ideality assumption
(see Sec.\ \ref{sec:melts}) becomes questionable
\cite{bryk_macdowell_08}.

In this chapter, I have chosen a field-theoretic way to present 
the SCF theory.  Freed and coworkers \cite{tang_freed_91, freed_95} 
have derived the same type of theory from a density functional
approach, using a reference system of non-interacting Gaussian
chains. Compared to the density functional approach, the
field-theoretic approach has the advantage that the effect of
fluctuations can be treated in a more transparent way (see Sec.\
\ref{sec:fluctuations}). On the other hand, information 
on the local liquid structure of the melt, ({\em i.e.}, monomer
correlation functions, packing effects etc.), can be incorporated
more easily in density functional approaches
\cite{woodward_91,woodward_yethiraj_94}. Density functionals have
also served as a starting point for the development of dynamical 
theories which allow to study the evolution of multiphase polymer 
blends in time \cite{fraaije_93, fraaije_vanvlimmeren_97, 
hasegawa_doi_97,kawakatsu_97} 
(see Sec.\ \ref{sec:dynamical_models_fields}).

\subsubsection{Fluctuation Effects}

\label{sec:fluctuations}

Mean-field approaches for polymer systems like the SCF theory tend
to be quite successful, because polymers overlap strongly and have
many interaction partners. However, there are several instances
where composition fluctuations become important and may affect the
phase behavior qualitatively.

\begin{figure}
\centerline{
\includegraphics[width=0.4\textwidth]{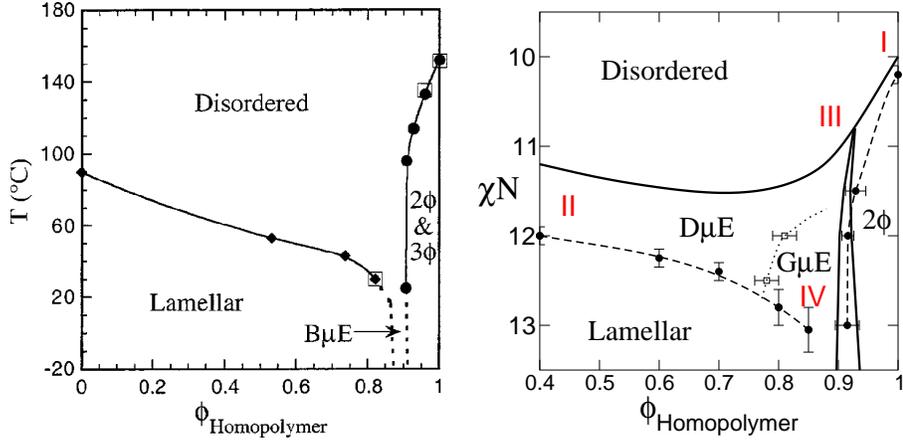}
\hspace{0.01\textwidth}
\includegraphics[width=0.39\textwidth,clip=true]{\dir/me_sim.eps}
} \caption{\label{fig:me_phases} Experimental phase diagram
for symmetric ternary blends of PDMS+PEE+PDMS-PEE ($N_C \sim 5 N_H$)
featuring lamellar phase, phase-separated region $2 \Phi$, and
microemulsion channel $B\mu E$ (left, from Ref.\ \protect
\citeonline{hillmyer_maurer_99}), compared with theoretical 
phase diagrams (right) from SCF theory (solid lines) and Monte Carlo 
simulations at $C=50$ (dashed lines/circles).
The dotted lines separate disordered regions with different local
structure: $D\mu E$, defect driven disorder, and $G\mu E$, genuine
microemulsion morphology. The regions I-IV are discussed in the text. 
Left Figure: Reprinted with permission from M.A. Hillmyer, W.W. Maurer,
T.P. Lodge, F.S. Bates, K. Almdal, J. Phys. Chemistry B {\bf 103}, 4814
(1999). Reproduced by permission of the Royal Society of Chemistry.
Right Figure: Adapted from Refs.\ \protect\citeonline{duechs_ganesan_03,
duechs_schmid_04} (D\"uchs et al 2003 and 2004).}
\end{figure}

To illustrate some of them, we show the phase diagram of ternary
mixtures containing A and B homopolymers and AB diblock copolymers
in Fig.\ \ref{fig:me_phases}. The left graph shows the experimental
phase diagram \cite{hillmyer_maurer_99}, the right graph theoretical 
phase diagrams obtained by D\"uchs {\em et al} \cite{duechs_ganesan_03,
duechs_schmid_04} from the SCF theory (solid lines) and from 
field-based computer simulations (dashed line, 
see Sec.\ \ref{sec:structural_models_fields} for details on the 
simulation method).  
Regions where different types of fluctuations come into play are 
marked by I -- IV.
\begin{itemize}
\item[I)]
Fluctuations are important in the close vicinity of {\em critical
points}, {\em i.e.}, continuous phase transitions. They affect the
values of the critical exponents, which characterize {\em e.g.}, the
behavior of the specific heat at the transition \cite{zinn_justin_book}.
In Fig.\ \ref{fig:me_phases}, such critical transitions are encountered
at high homopolymer concentration, where the system essentially behaves
like a binary A/B mixture with a critical demixing point. This point
belongs to the Ising universality class, hence the system should
exhibit Ising critical behavior. It has to be noted that in
polymer blends, critical exponents typically remain mean-field like until
very close to the critical point\cite{herktmaetzki_schelten_83,
schwahn_mortensen_87,holyst_vilgis_93}.
\item[II)]
The effect of fluctuations is more dramatic in the vicinity of
order-disorder transitions (ODT), {\em e.g.}, the transition between
the disordered phase and the lamellar phase at low homopolymer
concentrations. Fluctuations destroy the long-range order in weakly
segregated periodic structures, they shift the ODT and change the
order of the transition from continuous to first order (Brazovskii
mechanism \cite{brazovskii_75, fredrickson_helfand_87}). This effect
accounts for the differences between the experimental and the SCF
phase diagram in Fig.\ \ref{fig:diblock_1}.
\item[III)]
The SCF phase diagram features a three-phase (Lamellar + A + B)
coexistence region reaching up to a Lifshitz point. Lifshitz points
are generally believed to be destroyed by fluctuations in three
dimensions.
\item[IV)]
In strongly segregated mixtures, fluctuations affect the
large-scale structure of interfaces. Whereas mean-field interfaces
are flat, real interfaces undulate. The so-called 'capillary waves'
may destroy the orientational order in highly swollen lamellar
phases. A locally segregated, but globally disordered
'microemulsion' state intrudes between the homopolymer-poor lamellar
phase and the homopolymer-rich two-phase region in Fig.\
\ref{fig:me_phases}.
\end{itemize}
Both in the cases of III) and IV), the effect of fluctuations is to
destroy lamellar order in favor of a disordered state. However, the
mechanisms are different. This is found to leave a signature in the
structure of the disordered phase, which is still locally structured
with a characteristic wavevector $q^*$ \cite{duechs_schmid_04}. In the
Brazovskii regime, the wavevectore $q^*$ corresponds to that calculated
from the SCF theory (defect driven disorder regime, $D\mu E$). In the
capillary wave regime, the characteristic length scale increases,
compared to that calculated from the SCF theory (genuine
microemulsion regime, $G\mu E$).

Formally, the effect of fluctuations is hidden in the overall
prefactor $\rho_0/N$ in the SCF Hamiltonian ${\cal H}$ (Eq.\
\ref{eq:f_scf}). The larger this factor, the more accurate is the
saddle point integration that lies at the heart of the SCF
approximation. One can thus define a 'Ginzburg parameter' $C = R_g^d
\rho_0/N$, which characterizes the strength of the fluctuations.
Here the factor $R_g^d$ must be introduced to make $C$
dimensionless ($d$ is the spatial dimension), and $R_g = N b^2/6$ is
the natural length scale of the system, the radius of gyration of an
ideal chain of length $N$. The Ginzburg parameter roughly
corresponds to the ratio of the volume spanned by a chain, $R_g^d$,
and the volume actually occupied by a chain, $N/\rho_0$,
and thus measures the degree of interdigitation of chains. At $C \to
\infty$, the SCF approximation becomes exact. The numerical
simulations shown in Fig.\ \ref{fig:me_phases} were carried out at
$C=50$ (using the length of the copolymers as the reference length),
which still seems large. Nevertheless, the effect of fluctuations is
already quite dramatic.

In three dimensions ($d=3$), the Ginzburg parameter is proportional
to the square root of the chain length, $\sqrt{N}$. This is why the
mean-field theory becomes very good for systems of polymers with
high molecular weight, and only fails at selected points in the
phase diagram. The relation $C \propto \sqrt{N}$ has motivated the
definition of an 'invariant polymerization index' $\bar{N} = N b^6
\rho_0^2 \propto C^2$, which is also often used to quantify
fluctuation effects \cite{matsen_02,mueller_schmid_05}.
In two dimensional systems, $C$ is independent of the chain length
and fluctuation effects are much stronger. Furthermore,
topological constraints become important ({\em i.e.}, the fact that
chains cannot cross each other) which are not included in the
Helfand model. In three dimensional systems of linear polymers, 
they only affect the dynamics (leading to reptation), but in
two dimensions, they also change the static properties 
qualitatively \cite{semenov_03}.

Finally, we note that fluctuations can also be treated to some
extent within the SCF theory, by looking at Gaussian fluctuations
about the SCF solution\cite{shi_noolandi_96}. This is useful for
calculating structure factors and carrying out stability analyses.
However, Gaussian fluctuations alone cannot bring about the
qualitative changes in the phase behavior and the critical exponents
which have been described above.

\subsection{Analytical Theories}

The SCF equations have to be solved numerically, which can be quite
challenging from a computational point of view. In addition, they 
also serve as a starting point for the derivation of simpler
approximate theories, which may even have analytical solutions in
certain limits.

Two main regimes have to be distinguished here. In the {\em weak
segregation limit}, $\chi N$ is small, and the A and B homopolymers
or copolymer blocks are barely demixed. This is the realm of the
'random phase approximation' (RPA), which can be derived
systematically from the SCF theory. In the {\em strong segregation}
limit, $\chi N$ is large, the polymers or copolymer blocks are
strongly demixed and the system can basically be characterized in
terms of its internal interfaces.

\subsubsection{Weak Segregation and Random Phase Approximation}

\label{sec:wsl}

We first consider the situation at low $\chi N$. In this case, the
composition varies smoothly, A-rich domains still contain sizeable
fractions of B-monomers and vice versa, and the interfaces between
domains are broad, {\em i.e.}, their width is comparable to the
radius of gyration of the chains.

The idea of the RPA is to perform a systematic expansion about a
homogeneous reference state. More precisely, we use the SCF free
energy density functional, Eq. (\ref{eq:f_scf}), as a starting point, and
then expand ${\cal F}$ about the homogeneous state. Defining $\Phi_A
= \rho_A/\rho_0$ as usual, using $\Phi_A + \Phi_B \equiv 1$, and
introducing the Fourier representation $\Phi({\bf k}) = \int {\rm
d}{\bf r} \: e^{i {\bf kr}} \Phi({\bf r})$, we obtain a functional
of the form
\begin{eqnarray}
{\cal F}_{\mbox{\tiny RPA}}[\Phi_A({\bf r})] &= &
V \: \rho_0 \: f_{\mbox{\tiny homo}} + \kBT \:
\frac{\rho_0}{N} \:\Big\{ \:
\frac{1}{2V} \sum_{{\bf k}\ne 0}
|\Phi_A ({\bf k})|^2 \: \Gamma_2({\bf k})
\nonumber \\
&&
\frac{1}{6V^2} \sum_{{\bf k},{\bf k}'\ne 0}
\Phi_A ({\bf k}) \Phi_A({\bf k}') \Phi_A(-{\bf k} -{\bf k}')
 \: \Gamma_3({\bf k},{\bf k}')
+ \cdots
\Big\} ,
\label{eq:f_rpa}
\end{eqnarray}
where $f_{\mbox{\tiny homo}}$ is the SCF free energy per chain in
the reference system, and the coefficient $\Gamma_n$ depend on the
direct monomer interactions and on the intrachain correlations of
free ideal Gaussian chains.

We focus on the leading coefficient. To calculate $\Gamma_2$ for a
given blend, we define the pair correlators $K_{ij}$
\cite{shi_noolandi_96,schmid_98}
\begin{equation}
\label{eq:pair_correlator}
K_{ij}({\bf k}) =
\langle \hat{\rho}_i({\bf k}) \hat{\rho}_j({\bf k}) \rangle
\: \frac{1}{\rho_0 N},
\end{equation}
which give the density-density correlations in an identical blend of
noninteracting, ideal Gaussian chains and can thus be expressed in
terms of Debye functions $\gD(x)$ (Eq.\ (\ref{eq:debye})). For
example, for binary blends of homopolymers with chain length $N_j$,
gyration radii $R_{g,j}$, and mean volume fractions $\bar{\Phi}_j$
($j$=A,B), the pair correlators are given by
\begin{equation}\label{eq:k_homo}
K_{AA}=\frac{N}{N_A} \bar{\Phi}_A \gD(k^2R_{g,A}^2), \quad
K_{BB}=\frac{N}{N_B} \bar{\Phi}_B \gD(k^2 R_{g,B}^2), \quad
K_{AB}=K_{BA} = 0
\end{equation}
For pure diblock copolymer blends with A-fraction $f$, one gets
\begin{equation}
\label{eq:k_diblock}
\begin{array}{c}
K_{AA}=f^2\gD(f\: k^2 R_g^2)  \qquad
K_{BB}=(1-f)^2\gD((1-f)\:k^2R_g^2)
\\
K_{AB} = \frac{1}{2}  \Big( \gD(k^2 R_g^2) - f^2 \gD(f k^2 R_g^2) -
(1-f)^2 \gD( (1-f)k^2 R_g^2) \Big).
\end{array}
\end{equation}
Having calculated $K_{ij}$, one can evaluate $\Gamma_2$ according to
\cite{shi_noolandi_96,schmid_98}
\begin{equation}
\label{eq:gamma_2}
\Gamma_2 = \Big(
\frac{K_{AA} + K_{BB} + K_{AB} + K_{BA}}
{K_{AA} K_{BB} - K_{AB} K_{BA}}
- 2 \chi N
\Big).
\end{equation}
The function $\Gamma_2$ is particularly interesting, because it is
directly related to the structure factor of the homogeneous phase,
$S({\bf k}) \propto \Gamma_2({\bf k})^{-1}$. Hence the RPA provides
expressions for structure factors which can be compared to small
angle scattering experiments, {\em e.g.}, to determine effective
interaction parameters. This is probably its most important application.

We will now discuss specifically the application of the RPA to
binary homopolymer blends and to diblock copolymer blends.

\paragraph{(i) Binary homopolymer blends and
Flory-Huggins-de Gennes functional}~\\ 
\nopagebreak
\vspace{-0.8\baselineskip}

According to Eqs.\ (\ref{eq:k_homo}) and (\ref{eq:gamma_2}), the RPA
coefficient $\Gamma_2$ for binary blends is given by
\begin{equation}
\label{eq:G2_binary} \Gamma_2({\bf k}) = \Big(
\frac{N/N_A}{\bar{\Phi}_A \: \gD(k^2 R_{g,A}^2)}) +
\frac{N/N_B}{\bar{\Phi}_B \: \gD(k^2 R_{g,B}^2)}) -2 \chi N \Big).
\end{equation}
We assume that the composition varies only slowly in the system (on
length scales not much shorter than $R_g$), and expand $\Gamma_2$
for small wave vectors. Using $\gD(x) \approx 1 -x/3$ and $R_{g,j}^2
= b_j^2 N_j/6$, and inserting our result in the RPA expansion
(\ref{eq:f_rpa}), we obtain the free energy functional
\begin{eqnarray}
\nonumber
{\cal F}_{\mbox{\tiny RPA}}[\Phi_A] & \approx&
V \: \rho_0 \: f_{\mbox{\tiny homo}} + \kBT \: \rho_0
\:\Big\{ \:
\frac{1}{2V} \sum_{{\bf k}\ne 0}
(\frac{1}{\bar{\Phi}_A N_A} + \frac{1}{\bar{\Phi}_B N_B} - 2 \chi )
|\Phi_A({\bf k})|^2
\\ &&
\qquad + \:
\frac{1}{2V} \sum_{{\bf k}\ne 0}
\frac{1}{18} (\frac{b_A^2}{\bar{\Phi}_A} + \frac{b_B^2}{\bar{\Phi}_B} )
|\Phi_A({\bf k})|^2 \: k^2 \Big\}.
\label{eq:homo_rpa}
\end{eqnarray}
The first two terms in (\ref{eq:homo_rpa}) correspond to the second
order expansion of the integral $\rho_0 \int {\rm d}{\bf r} \:
f_{\mbox{\tiny SCF}}(\Phi_A)$, where $f_{\mbox{\tiny SCF}}(\Phi_A)$
is the SCF free energy per chain in a homogeneous system with
A-volume fraction $\Phi_A$. It thus seems reasonable to replace them
by the full integral. The last term is a square gradient
term in real space. Together, one recovers the Flory-Huggins-de
Gennes functional of Sec.\ \ref{sec:flory_huggins_degennes}, Eq.\
(\ref{eq:flory_huggins_degennes}).

\paragraph{(ii) Copolymer melts, Leibler theory, 
and Ohta-Kawasaki functional}~\\
\nopagebreak
\vspace{-0.8\baselineskip}

\label{sec:RPA}
In diblock copolymer blends, Eqs.\ (\ref{eq:k_diblock}) and
(\ref{eq:gamma_2}) yield the RPA coefficient
\begin{equation}
\label{eq:G2_diblock}
\Gamma_2({\bf k}) = \Big(
\frac{G(1)}{G(f)G(1-f) - (G(1)-G(f)-G(1-f))^2/4} - 2 \chi N
\Big)
\end{equation}
with the short hand notation $G(f) = f^2 \gD(f k^2 R_g^2)$. At low
$\chi N$, $\Gamma_2({\bf k})$ is positive. Upon increasing $\chi N$,
one encounters a spinodal line where $\Gamma_2({\bf k})$ becomes
zero for some nonzero $k= q^*$, and the disordered state becomes
unstable with respect to an ordered microphase separated state.

Since the function $\Gamma_2({\bf k})$ is spherically symmetric in
${\bf k}$, it does not favor a specific type of order. The
information on possible ordered states is contained in the higher
order coefficients $\Gamma_n$, most notably, in the structure of the
cubic term, $\Gamma_3$. In a seminal paper of 1979, Leibler
has carried out a fourth order RPA expansion and deduced a phase 
diagram which already included the three copolymer phases L, C, 
and S (Fig.\ \ref{fig:structures_cops}) \cite{leibler_80}. Milner and
Olmsted later showed that the Leibler theory is also capable of
reproducing the gyroid phase \cite{milner_olmsted_97}. The RPA phase
diagram roughly coincides with the full SCF phase diagram, as
established 1996 by Matsen and Bates \cite{matsen_bates_96}, up to
$\chi N < 12$. Unfortunately, fluctuations have a massive effect on
the phase diagram at these small $\chi N$ (see Fig.\ \ref{fig:diblock_1}),
therefore the predictive power of the Leibler theory must be
questioned. Nevertheless, it is useful for identifying potential
ordered phases and phase transitions in copolymer systems.
Generalized Leibler theories still prove to be efficient tools to
analyze phase transitions in complex copolymer blends by analytical
considerations \cite{erukhimovich_05}.

Next we attempt to construct a simplified free energy functional for
diblock copolymer melts, in the spirit of the Flory-Huggins-de
Gennes functional. To this end, we again expand $\Gamma_2$ in powers
of ${\bf k}$, as in (i). Compared to homopolymer blends, however,
there is an important difference: $\Gamma_2({\bf k})$ has a
singularity at ${\bf k} \to 0$ and diverges according to
\begin{equation}
\label{eq:G2_singularity}
\Gamma_2({\bf k}) \approx \frac{3}{2 f^2 (1-f)^2}
\: \frac{1}{k^2 R_g^2}
\qquad \mbox{at} \qquad k \to 0
.
\end{equation}
The singularity accounts for the fact that large-scale composition
fluctuations are not possible in copolymer blends, since the A- and
B- blocks are permanently linked to each other. It ensures that the 
structure factor $S({\bf k})$, vanishes at ${\bf k} \to 0$, 
suppresses macrophase separation and is thus ultimately
responsible for the onset of microphase separation in the RPA
theory.

A $1/k^2$ term like (\ref{eq:G2_singularity}) in a density functional 
corresponds to a long-range Coulomb type interaction. This observation 
motivated Ohta and Kawasaki \cite{ohta_kawasaki_86} in 1986 to propose 
a free energy functional for copolymer melts, which combines a regular
square-gradient functional accounting for direct short-range
interactions with a long-range Coulomb term accounting for the
connectivity of the copolymers. In real space, the Ohta-Kawasaki
functional has the form
\begin{equation}
\label{eq:ohta_kawasaki}
 {\cal F}_{\mbox{\tiny OK}}[\Phi_A] =
   \frac{\rho_0}{N} \int \!\! {\rm d}{\bf r} \:
   \Big\{
    {\cal W}(\Phi_A) + \frac{B}{2} (\nabla \Phi_A)^2
   \Big\}
  + \frac{\rho_0}{N} \:
   \frac{A}{2} \int \!\! {\rm d}{\bf r} \: {\rm d}{\bf r}' \:
   {\cal G}({\bf r},{\bf r'}) 
   \: \delta \Phi_A({\bf r}) \: \delta \Phi_A({\bf r}')
\end{equation}
with $\delta \Phi_A({\bf r}) = \Phi_A({\bf r}) - \bar{\Phi}_A$.
The last term introduces the long-range interactions,
with $G({\bf r},{\bf r}')$ defined such that
\begin{equation}
\label{eq:ohta_kawasaki_g}
\Delta {\cal G}({\bf r},{\bf r}') = - \delta({\bf r} - {\bf r}'),
\end{equation}
which corresponds to
${\cal G} ({\bf r},{\bf r}') \sim 1/|{\bf r}-{\bf r}'|$  in
infinitely extended systems.

Given Eq.\ (\ref{eq:G2_singularity}), it seems natural to identify
$A = 3/(2 f^2 (1-f)^2 R_g^2)$. The choice of ${\cal W}$ and $B$ is
somewhat more arbitrary. The function ${\cal W}(\Phi_A)$ is a free
energy density with two degenerate minima and can be approximated by
a fourth order polynomial in $\Phi_A$. As for the coefficient of the
square gradient term, $B$, Ohta and Kawasaki originally estimated it
from the asymptotic behavior of $\Gamma_2$ at $k \to \infty$,
\begin{equation}
\label{eq:G2_kinf}
\Gamma_2({\bf k}) \approx \frac{1}{2 f (1-f)}
\: {k^2 R_g^2}
\qquad \mbox{at} \qquad k \to \infty,
\end{equation}
which yields $B = R_g^2/(2 f (1-f))$. Later, they noted that this
choice of $B$ gives the wrong interfacial width at stronger
segregation, which has implications for the elastic constants and
the equilibrium period of the ordered phases, and suggested to
replace $B$ by a constant in the strong segregation limit
\cite{ohta_kawasaki_90}.

The Ohta-Kawasaki functional reproduces microphase separation and
complex copolymer phases such as the gyroid phase
\cite{uneyama_doi_05,yamada_nonomura_06} and even the Fddd phase
\cite{yamada_nonomura_06}. It can be handled much more easily than
the Leibler theory or the full SCF theory (see Sec.\
\ref{sec:dynamical_models_fields}), therefore it is
particularly popular in large-scale dynamical simulations of
copolymer melts (see Sec.\ \ref{sec:dynamical_models_fields}).
Different authors have generalized it to ternary blends containing
copolymers \cite{kawakatsu_94,kielhorn_muthukumar_99,uneyama_doi_05}.
In particular, Uneyama and Doi have recently proposed a general
density functional for polymer/copolymer blends that reduces to the
Flory-Huggins-de Gennes functional in the homopolymer case and to the
Ohta-Kawasaki functional in the diblock case.

\subsubsection{Strong Segregation}

\label{sec:ssl}

I turn to discussing the situation at high  $\chi N$. The A-rich
and B-rich (micro)phases are then well-separated by sharp
interfaces. The free energy contribution from the interfacial
regions (i) and the chain conformations inside the A- or B- domains
(ii) can be treated separately.

\paragraph{(i) Interfacial profiles and ground state dominance}~ \\
\nopagebreak
\vspace{-0.8\baselineskip}

In the interfacial region, the free energy is dominated by the
contribution of the direct A-B interactions and the local stretching
of segments. Chain end effects can be neglected. This simplifies the
situation considerably.

We first note that the diffusion equation ((\ref{eq:q_diffusion}) or
(\ref{eq:qdag_diffusion})) for a $j$-chain or a $j$-block of a chain
has the same structure than the time-dependent Schr\"odinger equation,
if one identifies $s \leftrightarrow it$. As is well known from quantum
mechanics, the general solution can formally be expressed as
\cite{sakurai_book} $q_j({\bf r},s) = \sum_n c_n \psi_{n,j}({\bf r})
e^{- \epsilon_{n,j} s}$, where $\{ \psi_{n,j}({\bf r}), \epsilon_{n,j} \}$
are Eigenfunctions and Eigenvalues of the operator $(b_j^2/6 \:
\Delta - W_j({\bf r})/N)$. At large $s$, the smallest Eigenvalue
$\epsilon_{0,j}$ dominates, {\em i.e.}, $q_j({\bf r},s) \propto
\psi_{0,j}({\bf r}) e^{- \epsilon_{0,j} s}$, and the resulting
density in the large $N_j$ limit is $\rho_j \propto |\psi_{0,j}|^2
e^{-\epsilon_{0,j} N_j}$. This type of approximation is called
'ground state dominance'. It is commonly used to study polymers at
interfaces and surfaces.

In the case of blends, we have the freedom to shift the fields
$W_j({\bf r})$ by a constant value, hence we can set $\epsilon_{0,j}
= 0$. The self-consistent field equations can thus be written as
\begin{equation}
\label{eq:ssl}
\rho_j = \rho_0 \: |\psi_j({\bf r})|^2
\quad \mbox{with} \quad
(\frac{b_j^2}{6} \Delta - \frac{W_j}{N}) \: \psi_j = 0
\quad \mbox{and} \quad
W_j = \frac{N}{\kBT}\:  \frac{\delta{\cal H}_I}{\delta \rho_j} - \xi,
\end{equation}
where $\psi_j$ is normalized such that ${\cal Q}_j = \int {\rm
d}{\bf r} \: |\psi_j({\bf r}|^2 = V_j$ is the partial volume
occupied by the polymers $j$, and $\xi ({\bf r})$ ensures $\sum_j
|\psi_j|^2 \equiv 1$.

In order to derive an epression for the free energy, we first note
that Eqs.\ (\ref{eq:ssl}) minimize a Lagrange action,
\begin{equation}
{\cal L} = {\cal H}_I + \frac{\kBT}{6} \: \rho_0 \: \int {\rm d}{\bf
r} \: ( b_A^2 (\nabla \psi_A)^2 + b_B^2 (\nabla \psi_B)^2 ),
\end{equation}
with respect to $\psi_j$ under the constraint $|\psi_A|^2 +
|\psi_B|^2 \equiv 1$. One easily checks that ${\cal L}$ vanishes for
homogeneous bulk states, and that the minimized ${\cal L}$ is equal
to the extremized SCF Hamiltonian ${\cal F}_{\mbox{\tiny SCF}}$,
Eq.\ (\ref{eq:f_scf}) up to a constant. Hence ${\cal L}$ can be
identified with the interfacial free energy. Rewriting it in terms
of the volume fractions $\phi_j$ and using $(\nabla \Phi_A)^2 =
(\nabla \Phi_B)^2$, one obtains the free energy functional
\begin{equation}
\label{eq:energy_ssl}
{\cal F}_{\mbox{\tiny int}}[\Phi_A({\bf r})] =
{\cal H}_I + \rho_0 \int {\rm d}{\bf r} \:
\: \frac{\kBT}{24}
( \frac{b_A^2}{\Phi_A} + \frac{b_B^2}{\Phi_B})  (\nabla \Phi_A)^2,
\end{equation}
which reproduces Eq.\ (\ref{eq:flory_huggins_ssl}) for
Flory-Huggins interactions (\ref{eq:hi_binary}).

For $b_A = b_B = b$ and Flory-Huggins interactions, the
self-consistent field equations (\ref{eq:ssl}) are solved by a
$\tanh$ profile, $(\rho_A - \rho_B) \sim \rho_0
\tanh(z/w_{\mbox{\tiny SSL}})$ with the interfacial width
$w_{\mbox{\tiny SSL}} = b/\sqrt{6 \chi}$ and the interfacial tension
$\sigma_{\mbox{\tiny SSL}} = \kBT \rho_0 b \sqrt{\chi/6}$
\cite{helfand_tagami_71,helfand_tagami_72,helfand_tagami_72b}.

\paragraph{(ii) Copolymer conformations and strong stretching theory}~\\
\nopagebreak
\vspace{-0.8\baselineskip}

The free energy functional (\ref{eq:energy_ssl}) is sufficient to
describe strongly segregated homopolymer blends. In copolymer
blends, additional contributions come into play due to the fact that the
copolymer junctions are confined to the interfaces and the copolymer
blocks stretch away from them into their respective A or B domains.
The associated costs of configurational free energy can be estimated
within a second approximation scheme, the 'strong stretching' theory
(SST) \cite{semenov_85,milner_witten_88,zhulina_borisov_90,
zhulina_borisov_91}.

The main idea of the SST was put forward in 1985 by
Semenov\cite{semenov_85}, who noted that for strongly stretched copolymer
blocks, the paths fluctuate around a set of 'most probable paths'.
This motivates to approximate the single-chain partition function
${\cal Q}$, Eq.\ (\ref{eq:part_q}), by its saddle point, {\em i.e.},
the path integral in ${\cal Q}$ is replaced by an integral over
'classical' paths ${\bf R}_c$ that extremize the integrand and thus
satisfy the differential equation \cite{milner_witten_88b}
\begin{equation}\label{eq:dgl_sst}
\frac{3}{b^2} \: \frac{{\rm d}^2 {\bf R}_c }{{\rm
d} s^2} = \frac{1}{N} \: \nabla W({\bf R}).
\end{equation}
We will treat the copolymer blocks as independent chains of length
$M$. The classical paths corresponding to one block are then
characterized by their boundary conditions, ${\bf R}_c(0) = {\bf
r}_j$ and ${\bf R}_c(M) = {\bf r}_e$, where the junction ${\bf r}_j$
is confined to an interface and the free end ${\bf r}_e$ is
distributed everywhere in its domain.

Next, we note that for infinitely long blocks, the classical
paths must satisfy
\begin{equation}
\label{eq:bc_sst} {\rm d} {\bf R}_c/{\rm d}s |_{s=M} = 0 \qquad
\mbox{for} \quad M \to \infty
\end{equation}
at the free end. Mathematically speaking, they would not have a
well-defined end position otherwise. Physically speaking, the
'average' chain representing the classical path does not sustain
tension at the free end, which seems reasonable. In the following,
Eq.\ (\ref{eq:bc_sst}) is also imposed for finite (large) blocks as
an additional boundary condition. Eq.\ (\ref{eq:dgl_sst}) is then
overdetermined and can, in general, no longer be solved for
arbitrary end positions ${\bf r}_e$. To ensure that chain ends are
indeed free to move throughout the domain, the field $W({\bf r})$ 
must have a special shape. Specifically, near flat interfaces it 
must be parabolic as a function of the distance $z$ from the interface
\cite{milner_witten_88,zhulina_borisov_90},
\begin{equation}
\label{eq:potential_sst}
\frac{1}{N} W(z) = - \frac{3}{8} \: \frac{\pi^2}{b^2 M^2} z^2.
\end{equation}
This is one of the main results of the SST. It generally applies to
situations where strongly stretched polymers are attached to an
interface, {\em e.g.}, strongly segregated copolymer blocks
\cite{witten_leibler_90,matsen_02}, or polymer brushes in solvents
of arbitrary quality \cite{zhulina_borisov_91}. The SST field must
always have the form (\ref{eq:potential_sst}), and the remaining
task is to realize this by a suitable choice of the chain end
distribution $P({\bf r}_e)$. In the incompressible blend case,
$P({\bf r}_e)$ must be chosen such that the density in the domains
is constant, $\rho_0$.

Luckily, we do not have to evaluate $P({\bf r}_e)$ explicitly to
calculate the free energy. The SST field has another convenient
property: One can show that the stretching energy of classical paths
of fixed length $N$ in a field satisfying Eqs.\ (\ref{eq:dgl_sst})
and (\ref{eq:bc_sst}) is exactly equal to the negative field energy,
\begin{equation}
\int_0^N {\rm d}s\: \frac{3}{2 b^2}
\Big( \frac{{\rm d} {\bf R}_c}{{\rm d} s} \Big)^2
= - \frac{1}{N} \int_0^N {\rm d}s \: W({\bf R}_c(s)).
\end{equation}
Summing over all blocks in a domain, the total stretching energy is
thus given by
\begin{equation}
\label{eq:energy_stretch}
\frac{{\cal F}_{\mbox{\tiny stretch}}}{\kBT}
= - \frac{1}{N} \int {\rm d}{\rm r} \:
\rho({\bf r}) \: W({\bf r})
\approx \rho_0 \frac{3}{8} \: \frac{\pi^2}{b^2 M^2}
\int {\rm d}{\bf r} \: z^2,
\end{equation}
where the integral is over the volume of the domain, and $z$ denotes
the closest distance to an interface. The total energy of the system
can be estimated as the sum over the stretching energies in the
different domains, Eq.\ (\ref{eq:energy_stretch}), and the
interfacial energy, Eq. (\ref{eq:energy_ssl}), and then used to
evaluate the relative stability of different phases. In the strong
segregation limit, only the C, L, and S phase are found to be stable
\cite{matsen_02}, in agreement with SCF calculations at high $\chi N$.

The validity of the strong stretching theory seems to be restricted
to very large chains \cite{matsen_01}. This is presumably to a large
extent due to the requirement (\ref{eq:bc_sst}), which does not 
necessarily hold for classical paths of finite length. Netz and Schick
\cite{netz_schick_97, netz_schick_98} have shown that an unrestricted
'classical theory', which just builds on the saddle point integration
of ${\cal Q}$ and avoids using (\ref{eq:bc_sst}), gives results that
agree better with the SCF theory. However, the classical theory has to
be solved numerically, and the computational advantage over the full SCF
theory is not evident.

The SST has found numerous applications \cite{matsen_02} and has been
extended and improved in various respect. It provides an analytical
approach to analyzing multicomponent polymer blends in a segregation
regime where the SCF theory becomes increasingly cumbersome, due to
the necessity of handling narrow interfaces.

\subsection{An Application: Interfaces in Binary Blends}

\label{sec:application_carelli}

\begin{figure}
\centerline{
\includegraphics[width=0.48\textwidth]{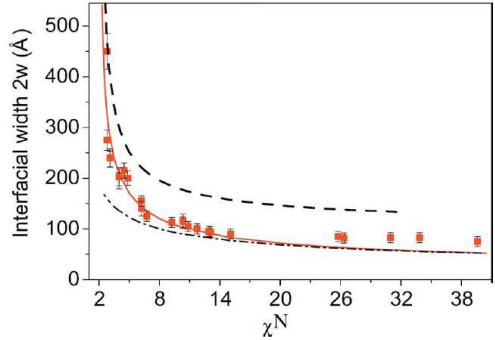}
}
\caption{\label{fig:width}
Intrinsic width of interfaces between A- and B-phases in
binary polyolefin blends as a function of $\chi N$,
compared with the predictions of the weak segregation theory
(long dashed line), the strong segregation theory (dot dashed
line), and the full SCF theory (solid line).
From Ref.\ \protect\citeonline{carelli_jones_05}.
}
\end{figure}

To close the theory section, I discuss the simplest possible
examples of an inhomogeneous polymer system: An interface in
a symmetrical binary homopolymer blend. This system has been
studied intensely in experiments \cite{fernandez_higgins_88,
stamm_schubert_95,schubert_stamm_96,kerle_klein_96,
sferrazza_xiao_97,carelli_jones_05b,carelli_jones_05}.
By mixing  random copolymers of ethylene and ethyl-ethylene with
two different, but very well defined copolymer ratios, Carelli
{\em et al.} \cite{carelli_jones_05b,carelli_jones_05}
were able to tune the Flory-Huggins parameter very finely and
study interfacial properties in a wide range of $\chi N$ between
the weak segregation limit and the strong segregation limit.
Fig.\ \ref{fig:width} compares the results for the interfacial
width and compares them with the mean-field prediction for the
weak segregation limit, the strong segregation limit, and the
full numerical result.

We should note that there is a complication here. As 
mentioned earlier, fluid-fluid interfaces are never flat, they
exhibit capillary waves \cite{buff_lovett_65,rowlinson_widom_book}.
This leads to an apparent broadening of the interfacial width $w$
\cite{semenov_94}. The apparent width depends on the lateral length
scale of the observation $L$ and can be calculated according to
\cite{lacasse_grest_98, werner_schmid_99}
\begin{equation}
\label{eq:width_broadening}
w^2 = w_0^2 + \frac{\kBT}{4 \sigma} \: \ln \Big( \frac{L}{a_0} \Big)
\quad \mbox{or} \quad
w^2 = w_0^2 + \frac{2 \kBT}{\pi \sigma} \: \ln \Big( \frac{L}{a_0} \Big),
\end{equation}
depending how it is measured. Here $w_0$ is the 'intrinsic' width,
$\sigma$ the interfacial tension, and $a_0$ a 'coarse-graining
length', which is roughly given by the interfacial width
\cite{semenov_94,werner_schmid_99}. Both the quantities $L$ and
$a_0$ are not very well defined in an actual experiment.
Fortunately, they only enter logarithmically, therefore the result
is not very sensitive to their values. The theoretical curves shown
in Fig.\ \ref{fig:width} include the capillary wave broadening,
calculated using the interfacial tension from the respective theory.

Comparing the curves in Fig.\ \ref{fig:width}, one finds that the 
weak segregation theory consistently overestimates the width, and the 
strong segregation theory consistently underestimates it. The numerical 
SCF values interpolate between the two regimes and are in excellent 
{\em quantitative} agreement with the experimental data over almost 
the whole range of $\chi N$.
The SCF theory is also found to perform very well compared to
computer simulations \cite{schmid_mueller_95, werner_schmid_99}. It
reproduces many features of the interfacial structure, such as chain
end distributions, local segment orientations etc.\ at a
quantitative level, if capillary waves are accounted for
\cite{werner_schmid_99}, This illustrates the power of the SCF
theory to describe the {\em local} structure of inhomogeneous
polymer systems, even if the {\em global} structure is affected by
large-scale composition fluctuations.

\section{Simulations of Multiphase Polymer Systems}

\label{sec:simulations}

Whereas theoretical work on multiphase polymer systems has a
long-standing tradition, the field of simulations in this area
is much younger. This is because polymer simulations are computationally
very expensive, which has essentially rendered them unfeasible until
roughly 20 years ago. In this section, I will attempt to give an
overview over the current state-of-the-art of simulations of
inhomogeneous multicomponent polymer systems.

\subsection{Coarse-Graining, Part II}

\label{sec:coarse-graining_2}

One of the obvious challenges in multiphase polymer simulations is
that polymers are such big molecules, which moreover self-organize
into even larger supramolecular structures. Polymeric materials
exhibit structure on a wide range of length scales, from the atomic
scale up to micrometers. Their specific material properties are to a
great extent determined by local inhomogeneities and internal
interfaces, and depend strongly on the interplay between these 
mesostructures in
space and time. In order to understand the materials and make
useful predictions for new substances, one must analyze their
properties on all time and length scales of interest. Therefore,
multiscale modeling has become one of the big topics in
computational polymer science.

The central element of multiscale modeling is coarse-graining. By
successively eliminating degrees of freedom (electronic structure,
atomic structure, molecular structure, etc.), a hierarchy of models is
constructed (see Sec.\ \ref{sec:coarse-graining_1}). For each type of
model, optimized simulation methods are developed, which allow one
to investigate specific aspects of the materials.

Having identified suitable classes of coarse-grained models,
one can proceed in two different manners:

\paragraph{Generic Modeling}~ \\
\label{sec:generic_coarse_graining}
\nopagebreak
\vspace{-0.8\baselineskip}

This approach has been favored historically, and up to date, the
overwhelming majority of simulations of multicomponent polymer
systems is still based on it (see Sec.\ \ref{sec:applications}). 
Generic models are simple and computationally efficient. They are 
not designed to represent specific materials; rather, they are the
simulation counterpart of the theoretical models discussed in the
previous section. They are suited to study generic properties of
polymer and copolymer systems, {\em i.e.}, to identify the behavior
that can be expected from their stringlike structure, their chemical
incompatibility etc. Simulations of generic models are also
particularly suited to test theories.

Generic models are used in all areas of materials science, and
in most cases, they only give qualitative insights into the behavior
of a material. This is different for polymers, because of their
universal properties (see Sec.\ \ref{sec:basic}). For example,
we have already seen in Fig.\ \ref{fig:width} that a generic
theoretical model (the Edwards model) quantitatively predicts
important aspects of the interfacial structure in real polymer
melts.

Nevertheless, the predictive power of generic models is
restricted, and relies on the knowledge of 'heuristic' parameters
such as the $\chi$-parameter. Therefore, a second approach 
is attracting growing interest.

\paragraph{Systematic Bottom-Up Modeling}~\\
\label{sec:systematic_coarse_graining}
\nopagebreak
\vspace{-0.8\baselineskip}

The idea of systematic coarse-graining is to establish a hierarchy
of models for the same specific material, starting from an ab-initio
description, with well-defined {\em quantitative} links between the
different levels.

Ideally, the goal is to replace many degrees of freedom by a
selection of fewer 'effective' degrees of freedom. If one is only
interested in equilibrium properties, the problem is at least
well-defined. For each possible coarse-grained configuration, one
must evaluate a partial partition function of the full system under
the constraints imposed by the values of the coarse-grained degrees
of freedom. This procedure results in an effective potential in the
coarse-grained space, which is, in general, a true multibody potential
-- it cannot be separated into contributions of pair potentials.
If one is interested in dynamical properties, the situation is even
more complicated. One must replace a dynamical system for all
variables by a lower dimensional system for a subset of effective
variables. This can be done approximately, {\em e.g.}, using Mori
Zwanzig projector operator techniques \cite{forster_book}. The new 
dynamical system is inevitably a stochastic process with memory, 
{\em i.e.}, the future time evolution not only depends on the current 
state of the system, but also on its entire history.

Obviously, such 'ideal coarse-graining' is not feasible for
polymer systems. Instead, researchers adopt a heuristic approach,
where they first define a coarse-grained model, which typically has
no memory and only pair potentials, and match the properties
of the coarse-grained model with those of the fine-grain model as
good as they can: The model parameters are chosen such that the
coarse-grained model reproduces physical properties of interest,
such as correlation functions or diffusion constants
\cite{baschnagel_binder_00, muellerplathe_02, muellerplathe_03,
binder_paul_04}.

Already early on, researchers have started to develop schemes for
mapping real polymers on lattice models \cite{baschnagel_binder_91,
paul_pistoor_94,rapold_mattice_95}. Nowadays, off-lattice models are 
more common. Early approaches focussed on the task of reproducing 
the correct {\em intrachain} correlations by optimizing the bond 
potentials in the chains \cite{baschnagel_binder_91}. Later, the
{\em interchain} correlations were considered as well, which can
be matched by adjusting the non-bonded, intermolecular potentials
in the coarse-grained model. It is important to note that the resulting
effective potentials depend on the concentration and the temperature
\cite{carbone_varzaneh_08} (much like the $\chi$-parameter itself).
Different methods to determine effective potentials have been
devised and even automated packages are available
\cite{muellerplathe_03,reith_puetz_03, faller_04b,
milano_muellerplathe_05,li_kou_06}. The reverse problem -- how to
reconstruct a fine-scale model from a given coarse-scale
configuration -- has also been addressed \cite{santangelo_matteo_07,
sewell_rasmussen_07}. Nowadays, the available techniques for mapping
static properties are relatively advanced. In contrast, the field of
mapping dynamical properties is still in its infancy
\cite{guenza_08}.

The standard multiscale approach is sequential, {\em i.e.},
numerical simulations are carried out separately for different
levels. Currently, increasing effort is devoted to developing hybrid
schemes where several coarse-graining levels are considered
simultaneously within one single simulation
\cite{prapotnik_dellesite_08, honda_kawakatsu_07}.

Despite the large amount of work that has already been devoted to
systematic coarse-graining, coarse-grained simulation studies of
realistic multicomponent polymer blends are still scarce. Mattice
and coworkers have carried out lattice simulations of blends
containing polyethylene and polypropylene homopolymers and
copolymers \cite{akten_mattice_01,rane_mattice_04,choi_rane_06}.
Faller {\em et al} have developed and studied a coarse-grained model
for blends of polyisoprene and polystyrene
\cite{sun_faller_06,sun_faller_07,sun_pon_07}.

\subsection{Overview of structural models}

\label{sec:structural_models}

After these general remarks, I shall give a brief overview over the
different models that are currently used in multicomponent polymer
simulations.

\subsubsection{Atomistic Models}

\label{sec:atomistic}

Atomistic simulations are computationally intensive, and rely very
much on the quality of the force fields. (Force fields are a
separate issue in multiscale modeling, which shall not be discussed
here.) Therefore, atomistic simulations of blends are
still relatively scarce. So far, most studies have focussed on
miscibility aspects \cite{ choi_jo_97, lee_lee_99,
gestoso_brisson_03, gestoso_brisson_03b, maranas_kumar_98,
maranas_mondello_98, neelakantan_maranas_04, faller_04,
rissanou_peristeras_07,prathab_subramanian_07, prathab_aminabhavi_07}.
Already early on, atomistic and mesoscopic simulations were combined
in multiscale studies: Atomistic simulations were used to determine
the Flory-Huggins $\chi$-parameter, coarse-grained methods were then
applied to study large-scale aspects of phase separation
\cite{tao_fan_94, spyriouni_vergelati_01, yang_li_05,
jawalkar_adoor_05, jawalkar_aminabhavi_06, jawalkar_raju_07,
jawalkar_nataraj_08} or mesophase formation \cite{ lee_ju_07}. Only
few fully atomistic studies deal with aspects beyond miscibility,
{\em e.g}, the formation of lamellar structures in diblock
copolymers \cite{lu_li_98}, or the diffusion of small molecules in
blends \cite{pavel_shanks_05}.

\subsubsection{Coarse-Grained Particle Models}

\label{sec:structural_models_particles}

The coarse-grained models for polymers can be divided into two main
classes: {\em Coarse-grained particle} models operate with
descriptions of the polymers that are considerably simplified,
compared to atomistic models, but still treat them as explicit
individual objects. {\em Field} models describe polymer systems in
terms of spatially varying continuous fields. I begin with
discussing some of the most common particle models.

\paragraph{Lattice Chain Models}~ \\
\nopagebreak
\vspace{-0.8\baselineskip}

Lattice models have the oldest tradition among the coarse-grained
particle models for polymer simulations, and are still very popular.
The first molecular simulations of multiphase polymer systems
-- studies of binary homopolymer blends by Sariban and Binder
in 1987\cite{sariban_binder_87, sariban_binder_88} and
by Cifra and coworkers in 1988 \cite{cifra_karasz_88} -- were
based on lattice models. They are particularly suited to be studied
with Monte Carlo methods, and several smart Monte Carlo algorithms
have been designed especially for lattice polymer simulations
\cite{frenkel_smit_book,erice_v2_book}.

In molecular lattice models, the polymers are represented as strings
of monomers confined to a lattice. A natural approach consists in
placing the 'monomers' on lattice sites and linking them by bonds
that connect nearest-neighbor sites. For many applications, 
it has proven useful to apply less rigid constraints on the links
and allow for bonds of variable length, which may also connect
second-nearest neighbors \cite{larson_scriven_85} or stretch over
even longer distances \cite{dotera_hatano_96}. Moreover, the lattice
is usually not entirely filled with monomers, but also contains a
small fraction of voids. This is because most Monte Carlo algorithms
for polymers do not work at full filling, and special
algorithms have to be devised for that case\cite{pakula_87}. One
particularly popular lattice model is the 'bond-fluctuation model',
devised in 1988 by Carmesin and Kremer \cite{carmesin_kremer_88}.
It is based on the cubic lattice; monomers do not occupy
single sites, but entire cubes in a cubic lattice. They are
connected by 'fluctuating bonds' of varying length, ($2,
\sqrt{5},\sqrt{6},3$ or $\sqrt{10}$ lattice constants). In the
bond-fluctuation model, a polymer system behaves like
a dense polymer melt already at the volume fraction $0.5$.
Therefore, it can be simulated very efficiently.

Despite their intrinsically anisotropic character, lattice models
are able to reproduce most known self-assembled mesophases in
copolymer melts, even the gyroid phase in diblock copolymers
\cite{martinez_escobedo_05}. Nowadays, they are used to study such
complex systems as ABC triblock copolymer melts confined in
cylindrical nanotubes \cite{feng_ruckenstein_07}, which feature a
rich spectrum of novel morphologies, {\em e.g.}, stacked disks,
curved lamellar structures, and various types of helices.

\paragraph{Off-Lattice Chain Models}~\\
\nopagebreak
\vspace{-0.8\baselineskip}

For many years, only lattice simulations were sufficiently efficient
that they could be used to study polymer blends at a molecular level.
With computers becoming more and more powerful, off-lattice chain
models become increasingly popular. Compared to lattice models, they
have the advantage that they provide easy access to forces and can
also be used in Molecular Dynamics or Brownian Dynamics simulations.
They do not impose restrictions on the size and shape of the
simulation box (in lattice models, the box dimensions have to be
multiple integers of the lattice constant). The structure of space
is not anisotropic as in lattice models. Whereas the inherent anisotropy 
of lattice models does not seem to cause problems if the lattice model
is sufficiently flexible and if the chains are sufficiently long,
simulations of shorter chains can be hampered by lattice artifacts.

In bead-spring models, polymers are represented by chains of 
spherically symmetric force centers connected by springs. Numerous 
variants have been proposed, which differ in the choice of the spring 
potentials (the bonded interactions) and the choice of the pairwise 
interactions between the beads (the non-bonded potentials).

The simplest choice of spring potential is a harmonic potential. In
order to prevent chains from crossing each other in dynamical
simulations, an anharmonic cutoff on the spring length is often
imposed. A popular choice is the 'Finitely Extensible Nonlinear
Elastic' (FENE) potential
\begin{equation}
 V_{\mbox{\tiny FENE}}(b) = \frac{k}{2} \:
   d^2 \: \ln\Big( 1 - \frac{(b-b_0)^2}{d^2} \: \Big),
\end{equation}
which reduces to a harmonic spring
potential with equilibrium spring length $b_0$ at $b \approx b_0$,
and diverges at $|b-b_0| \to d$. In some applications, the springs
are constrained to have fixed lengths -- however, this requires the
use of special algorithms and changes slightly the distribution of
bond angles \cite{frenkel_smit_book}. In addition, some bead-spring
models also include bending potentials that allow one to tune the
chain stiffness, or even torsional potentials.

The non-bonded interactions drive the segregation of the monomers.
As we have discussed in Sec.\ \ref{sec:flory_huggins_parameter},
both energetic and entropic factors can contribute to making
monomers incompatible. Many models operate with energetic monomer
(in)compatibi\-li\-ties, but models with entropically driven segregation
are also common. As an example, we consider one commonly used type
of potential, the truncated Lennard-Jones potentials
\begin{equation}
 V_{\mbox{\tiny LJ}}(r) = \epsilon
  \Big( (\sigma/r)^{12} - (\sigma/r)^6 + C \Big)
  \quad \mbox{for} \quad r < r_c
\end{equation}
($V_{\mbox{\tiny LJ}}=0$ otherwise), where the parameter $C$ is
chosen such that $V_{\mbox{\tiny LJ}}(r)$ is continuous at $r=r_c$.
If the cutoff parameter $r_c$ is larger than $2^{1/6}\sigma$ (a
common choice is $r = 2.5 \sigma$), the potential has a repulsive
core surrounded by an attractive well . In this case, energetic
incompatibility can be imposed by using species dependent
interaction parameters $\epsilon_{ij}$ with
$2\epsilon_{AB}<\epsilon_{AA}+\epsilon_{BB}$. If the cutoff
parameter is $r_c \le 2^{1/6}$, the potential is purely repulsive.
In this case, one can still enforce monomer segregation by choosing
species dependent and {\em non-additive} interaction radii
$\sigma_{ij}$ with $2\sigma_{AB}>\sigma_{AA}+\sigma_{BB}$. The
mechanism driving the segregation is the Equation-of-State effect
discussed in Sec.\ \ref{sec:flory_huggins_parameter}. A simulation
model that is based on this idea has been proposed by Grest, Lacasse
and Kremer in 1996 \cite{grest_lacasse_96}. It is more efficient
than conventional models with energetically driven segregation,
because the interaction range is shorter. We note that high
pressures have to be applied to keep the chains together and to
drive the demixing.

In the context of 'Dissipative particle dynamics' (DPD) simulations,
(see below) it has also become popular to use soft non-bonded
potentials without a hard core. A typical DPD-potential has the
form \cite{groot_warren_97}
\begin{equation}\label{eq:v_dpd}
 V_{\mbox{\tiny DPD}}(r)
 = \frac{v}{2} \: \Big( 1-r/r_c \Big)^2 \quad \mbox{for}\quad r < r_c
\end{equation}
($V_{\mbox{\tiny DPD}}=0$ otherwise). Demixing is induced by using
species dependent parameters $v_{ij}>0$ with $2v_{AB} > (v_{AA}+ v_{BB})$.
The mechanism driving the segregation is again related to the
Equation-of-State effect -- like particles overlap more strongly than
unlike particles. The DPD simulation method was originally proposed in
the context of fluid simulations, where every DPD particle supposedly
represents a lump of true particles. This motivates the absence of hard,
impenetrable cores and even a roughly linear shape \cite{klapp_diestler_04}.
Early on, DPD potentials were also used in polymer simulations
\cite{groot_warren_97}. As long as topological constraints are not
important, the monomer potentials do not need to have a hard core
(see also the discussion at the end of the next section).

\paragraph{'Edwards' Models}~\\
\nopagebreak
\vspace{-0.8\baselineskip}

A special class of chain models are the Edwards models, which shall
be treated separately. The idea is to implement directly an 
Edwards-type interaction ${\cal H}_I$ (Eq.\ (\ref{eq:hi})) in a
particle-based simulation. The molecules are modeled as off-lattice
chains as before, and the non-bonded interactions are given by a
potential $V = {\cal H}_I[\{\rho_i\}]$ that depends on local monomer
densities $\rho_i({\bf r})$. To complete the
definition of the model, one must prescribe how to evaluate the
local densities. This is most easily done by simply counting the
monomers in each cell of a grid. Other prescriptions that do not
impose a grid are also conceivable.

Edwards models are not yet very common in simulations of multiphase
polymer systems, but they will very likely gain importance in the
future. The basic idea was put forward by Zuckermann and coworkers
\cite{laradji_guo_94, soga_zuckermann_96} in 1994 in the context of
polymer brush simulations. It was first applied to studying
microphase separation in block copolymers by Besold {\em et al.}
\cite{besold_hassager_99}, who also showed that the model produces
correct single-chain behavior in solution \cite{besold_guo_00} ({\em
i.e.}, the chains behave like regular self-avoiding chains). In the
following, the power of the approach has been demonstrated in
a series of impressive work by M\"uller and
coworkers\cite{detcheverry_kang_08} (see also Sec.\
\ref{sec:dynamical_models_particles}).

Two notes are in order here. First, it is difficult to impose strict
incompressibility in particle-based simulations. Instead, dense melt
simulations usually operate at finite compressibility: One
introduces an additional term \cite{helfand_tagami_71}
\begin{equation}
{\cal H}_{I,\mbox{\tiny comp.}} = \frac{\kappa}{2} (\rho_A+ \rho_B -
\rho_0)^2
\end{equation}
with a high modulus $\kappa$ in the interaction Hamiltonian. Second,
chains may overlap in the Edwards models. They share this 'problem'
with the DPD models introduced in the previous section. In three 
dimensions, the lack of hard core interactions has no effect on the 
static properties of linear polymers \cite{besold_guo_00,
detcheverry_kang_08}. Topological constraints become important in
two dimensions \cite{semenov_03}, or in melts of cyclic polymers
\cite{mueller_wittmer_96}, or, most notably, when looking at dynamical
properties. In Edwards models, chains do not entangle, and reptation
dynamics has to be put in 'by hand' \cite{mueller_daoulas_08b}.

\paragraph{Ellipsoid Model}~\\
\nopagebreak
\vspace{-0.8\baselineskip}

In 1998, Murat and Kremer proposed a model that allows to study
weakly segregated polymer blends on the scale of the gyration radius
$R_g$ and beyond \cite{murat_kremer_98}. If details of the
conformations are not of interest, the polymers can be replaced by
single, soft particles with ellipsoidal shape. The idea was then
pursued mainly by Eurich, Maass, and coworkers
\cite{eurich_maass_01,eurich_maass_02,eurich_karatchentsev_07}, who
also suggested to extend the model to diblock copolymers by modeling
them as dimers to account for their dumbbell shape
\cite{eurich_karatchentsev_07}. Sambrisky and Guenza have recently
worked out a microscopic foundation for coarse-graining diblock 
copolymers into such dumbbells, which is based on liquid-state 
integral equations \cite{sambriski_guenza_07}.

\subsubsection{Field-Based Models}

\label{sec:structural_models_fields}

Field-based models for polymer systems no longer treat the molecules
as individual objects, but describe them in terms of locally varying
fluctuating fields. Among these, the {\em molecular} field theories
still incorporate some information on the conformations of molecules,
and the {\em Ginzburg-Landau} models focus on the large-scale
structure only.

\paragraph{Field-Theoretical Models}~\\
\nopagebreak
\vspace{-0.8\baselineskip}

Field-theoretical models are the molecular field equivalent of the
particle-based 'Edwards' models discussed above. The idea is to use
the starting point of the self-consistent field theory, {\em i.e.},
a field-theoretic expression for the partition function
({\em e.g.}, Eq.\ (\ref{eq:partition_2})), and to evaluate the 
functional integrals over fluctuating fields by simulation methods.
Since some fields are complex, one is confronted with a sign problem
(the integrand oscillates between positive and negative).
Fredrickson and coworkers have demonstrated that the integrals can
nevertheless be evaluated in many cases using a method borrowed from
elementary particle physics, the 'Complex Langevin' simulation
method \cite{fredrickson_ganesan_02,
fredrickson_book,fredrickson_07}. We refer to Ref.\
\citeonline{fredrickson_book} for a general presentation
of the method, and Ref.\ \citeonline{lennon_mohler_08} for technical
details.

To be more specific, let us consider the system discussed in Sec.\
\ref{sec:scf_principle}, a blend of polymers/copolymers containing
two types of monomers A and B, with Flory-Huggins interactions
(\ref{eq:hi_binary}). Rather than evaluating the integral over all
five fluctuating fields $W_{A,B},\xi,\rho_{A,B}$ in Eq.\
(\ref{eq:partition_2}), we reconsider the original partition
function, Eq.\ (\ref{eq:partition_1}), and decouple the integrals
over different paths (polymer conformations) by means of a
Hubbard-Stratonovich transformation \cite{zinn_justin_book}. This is
possible, because the interaction Flory-Huggins interaction ${\cal
H}_I[\{\hat{\rho}_i\}]$ is quadratic in the densities
$\hat{\rho}_i$. The result is a functional integral over 'only' two
fields -- one which is conjugate to the total density $\hat{\rho} =
\hat{\rho}_A + \hat{\rho}_B$ and imaginary, and one which is
conjugate to the composition $\hat{m} = \hat{\rho}_B - \hat{\rho}_A$
and real:
\begin{equation} \label{eq:partition_3}
  {\cal Z} \propto
    \int_{i \infty} {\cal D} W_+ \int_{\infty} {\cal D} W_-
    \: e^{-{\cal F}_{\mbox{\tiny FTS}}/\kBT}
\end{equation}
with
\begin{equation}
{\cal F}_{\mbox{\tiny FTS}}[W_+,W_-] = \frac{\rho_0}{N} \Big\{
\frac{1}{\chi N} \int {\rm d}{\bf r} \: W_-^2 - \int {\rm d}{\bf r}
\: W_+ - \sum_\alpha V_\alpha \frac{N}{N_\alpha} \: \ln \big( \rho_0
{\cal Q}_\alpha/n_\alpha \big) \: \Big\}.
\end{equation}
This partition function is the starting point for field-theoretical
simulations of incompressible AB (co)polymer systems.

The idea to study multiphase polymer systems by direct evaluation of
fluctuating field integrals like Eq.\ (\ref{eq:partition_3}) was
first put forward by Ganesan and Fredrickson in 2001
\cite{ganesan_fredrickson_01}. They used Complex-Langevin
simulations to look at fluctuation effects in pure symmetric
diblocks. Since then, Fredrickson and coworkers have shown that the
method can be extended in various ways, {\em e.g.}, it can deal with
external stresses \cite{barrat_fredrickson_05}), and it can be
applied very naturally to charged polymers \cite{popov_lee_08,
lee_popov_08}. Nevertheless, there are still problems with it.
The theoretical foundations of the Complex Langevin method are 
not fully established. One can show that under certain conditions, 
it produces the correct statistical averages {\em if} the system 
reaches equilibrium \cite{fredrickson_book}. However, the stability 
of the simulations cannot be ensured. Even if the Langevin step size 
is chosen small, a fraction of simulation runs still crashes 
\cite{lennon_mohler_08}. Such problems are unknown from other 
simulation methods. Since it is generally not widely used and 
very young in the polymer community, the Complex Langevin method
still suffers from teething troubles.

An alternative approach to carrying out field-theoretical
simulations in melts has been suggested by D\"uchs {\em et. al.}
\cite{duechs_ganesan_03} (see also Ref.\ \citeonline{mueller_schmid_05}).
They suggested to treat only the real integral over $W_-$ with
computer simulation and approximate the problematic imaginary
integral over $W_+$ by its saddle-point. The integral over $W_-$
can be evaluated by standard Monte Carlo methods. The results from
this partial saddle point approach were shown to agree quantitatively
with the full Complex Langevin simulation \cite{duechs_ganesan_03}.
Possible strategies to improve upon the saddle point integral have 
been pointed out by B\"aurle and coworkers \cite{moreira_baeurle_03,
baeurle_efimov_06,baeurle_efimov_06b}.

As already pointed out, field-theoretic models are in some sense
equivalent to particle-based Edwards models (if the latter use
discrete Gaussian chain models). In field-theoretical simulations,
the mean-field limit can be reached very naturally and at low
computational cost. Hence they are more efficient than
particle-based simulations at high polymer densities, close to the
SCF limit, whereas particle-based simulations are superior at lower
polymer densities.

An application of field-theoretic simulations was already shown in
Sec.\ \ref{sec:fluctuations}. The simulation data shown in Fig.\
\ref{fig:me_phases} were obtained with field-theoretic Monte Carlo
simulations.

\paragraph{Molecular Density Functionals}~\\
\nopagebreak
\vspace{-0.8\baselineskip}

Molecular density functionals are free energy functionals of the
type (\ref{eq:f_scf}). 
They are used in dynamical density functional simulations (see Sec.\
\ref{sec:dynamical_models_fields}).

\paragraph{Ginzburg-Landau Models}~\\
\nopagebreak
\vspace{-0.8\baselineskip}

Ginzburg-Landau models no longer incorporate specific information
on chain conformations and thus have a much simpler structure than
molecular field models. Examples are the Flory-Huggins-de Gennes
functional for homopolymer blends
(Eq.\ (\ref{eq:flory_huggins_degennes})),
or the Ohta-Kawasaki functional for diblock copolymer melts
(Eq.\ (\ref{eq:ohta_kawasaki})). This is the highest level
of coarse-graining discussed in the present article.

\subsection{Overview of Dynamical Models}

\label{sec:dynamical_models}

In many cases, only static equilibrium properties are of interest,
and then most dynamical models are equivalent. When looking
at dynamical properties such as dynamical response functions, or at
nonequilibrium situations, the choice of the dynamical model becomes
relevant. In the following, I summarize some important models that
have been used to study multiphase polymer systems.

\subsubsection{Particle-Based Dynamics}

\label{sec:dynamical_models_particles}

\paragraph{Kinetic Monte Carlo (MC)}~\\
\nopagebreak
\vspace{-0.8\baselineskip}

{\em A priori}, the Monte Carlo (MC) method has been invented as a
method to evaluate high dimensional integrals ({\em i.e.}, thermal
averages), and is designed for studying dynamics.
Nevertheless, MC simulations are used for dynamical studies, based
on the fact that like many static properties, dynamical phenomena
are also often governed by universal principles. In kinetic MC
simulations, one analyzes the artificial Monte Carlo evolution of a
system in order to gain insight into real dynamical processes in the
system. The main requirement is that the Monte Carlo moves are only
local and reasonably 'realistic', {\em i.e.}, chain crossings are
not allowed. Kinetic Monte Carlo simulations have been used, {\em
e.g.}, to study the early stages of demixing in binary blends or the
ordering dynamics in block copolymer melts.

\paragraph{Molecular Dynamics (MD)}~ \\
\nopagebreak
\vspace{-0.8\baselineskip}

In Molecular Dynamics simulations, one solves directly Newton's or
Hamilton's equations of motions. This is the most straightforward
approach to studying dynamical processes in a system. As usual, of
course, the devil is in the details \cite{frenkel_smit_book}.

\paragraph{Brownian Dynamics}~\\
\nopagebreak
\vspace{-0.8\baselineskip}

In Brownian dynamics simulations, one also solves equations of
motion, but the underlying dynamics is not Hamiltonian. As we have
noted earlier (Sec.\ \ref{sec:systematic_coarse_graining}),
systematic coarse-graining necessarily turns a dynamical system 
into a stochastic process. Brownian Dynamics simulations account 
for this fact.  They include dissipative and stochastic forces, 
which supposedly represent the interaction of the coarse-grained 
degrees of freedom with those that have been integrated out. 
In general, however, these forces are not derived systematically 
from the original model, but postulated heuristically.

More specifically, the particles $j$ experience three types of
forces:
\begin{equation}
m_j \ddot{\bf r}_j = {\bf F}_j^C + {\bf F}_j^D + {\bf F}_j^R
\end{equation}
The first term ${\bf F}_j^C$ correspondes to the conservative forces, 
which are derived from the interparticle potentials of the structural
model under consideration. These forces also enter the standard
Molecular Dynamics simulations. The second ${\bf F}_j^D$ term is a
dissipative force that couples to the velocity of the particles. The
last term ${\bf F}_j^R$ is a Gaussian stochastic force with mean
zero whose amplitude is related to the dissipative force by a
'fluctuation-dissipation theorem' \cite{risken_book}. In a canonical
simulation, the last two terms constitute a thermostat, {\em i.e.}, they
maintain the system at a given temperature $T$. Microcanonical
models where energy is randomly shifted between 'internal' and
'external' degrees of freedom have been designed as well.

In the simplest (canonical) Ansatz, the dissipative force on a particle 
$j$ at time $t$ is proportional to its velocity ${\bf v}_j(t)$. According 
to the fluctuation-dissipation theorem, the stochastic force then fulfills
\begin{equation}\label{eq:langevin_dynamics}
 {\bf F}_j^D = - \gamma_j {\bf v}_j \qquad \longleftrightarrow \quad
 \langle F_{j,\alpha}^R(t) \: F_{k,\beta}^R (t')\rangle
 = 2 \gamma_j \: \kBT \: \delta_{\alpha \beta} \:
 \delta_{jk} \: \delta(t-t').
\end{equation}
Other choices are possible. For example, the dissipative force on
$j$ can depend on the velocities of other particles $k$ as well,
and/or on the history of the system. The fluctuation-dissipation
relation for the stochastic force has to be adjusted accordingly
\cite{risken_book}. The dynamics defined by
Eq.\ (\ref{eq:langevin_dynamics}) is commonly used
because it is so simple, but it has the disadvantage that it is not
Galilean-invariant -- a system moving at constant speed is treated
differently than a system at rest. Therefore, it does not
incorporate hydrodynamic effects correctly, and it is not suited to
study nonequilibrium systems such as polymer melts in shear flow.

\paragraph{Dissipative Particle Dynamics (DPD)}~\\
\nopagebreak
\vspace{-0.8\baselineskip}

The problems of Eq.\ (\ref{eq:langevin_dynamics}) are avoided in the
recently developed 'Dissipative Particle Dynamics' (DPD) method
\cite{hoogerbrugge_koelman_92,espanol_warren_95}. DPD is a special type
of Brownian Dynamics which is Galilean-invariant and has become very
popular in simulations of complex fluids in recent years. The dissipative
and stochastic forces are constructed such that they conserve the total
momentum and angular momentum in the system. Consequently, they
couple to relative velocities of particles rather than absolute
velocities, and they act as central forces (to preserve angular
momentum). Specifically, the forces acting on a particle $j$ are
given by
\begin{equation}
 {\bf F}_j^C- \sum_{k \ne j} \Big\{
 \gamma \: \omega(r_{jk}) \: (\hat{\bf r}_{jk} {\bf v}_{jk}) \:
 \hat{\bf r}_{jk}
 + \sqrt{2 \gamma \: \kBT \: \omega(r_{jk})} \:
 \hat{\bf r}_{jk} \: \zeta_{jk}
 \Big\},
\end{equation}
where ${\bf r}_{jk} = {\bf r}_j - {\bf r}_k$ is the vector
separating two particles $j$ and $k$, $r_{jk}$ its length, $\hat{\bf
r}_{jk}$ the unit vector in the same direction, $\omega(r)$ an
arbitrary function with a cutoff, and $\zeta_{jk}$ are symmetric and
uncorrelated Gaussian random numbers with mean zero and variance
one.

In the literature, the term 'DPD simulations' often refers to
simulations that use DPD dynamics in combination with soft 'DPD
potentials' (see Sec.\ \ref{sec:structural_models_particles},
(Eq.\ (\ref{eq:v_dpd})). However, DPD can of course be
used as a dynamical model \cite{espanol_warren_95} or simply as
a thermostat \cite{soddemann_duenweg_03} in combination with any
type of potential.

In contrast to kinetic Monte Carlo simulations or simple
Brownian dynamics simulations (using Eq.\ (\ref{eq:langevin_dynamics})),
DPD simulations take full account of hydrodynamic interactions.
Studies of microphase separation in copolymer melts have shown that
this makes a difference. The dynamics is strongly affected by
hydrodynamic effects in certain regions of phase space - in particular,
hydrodynamic interactions play a critical part in helping the
system to escape from metastable transient states \cite{groot_madden_99}.

\paragraph{Single Chain in Mean Field (SCMF)}~\\
\nopagebreak
\vspace{-0.8\baselineskip}

In 2005, M\"uller proposed an efficient method to study the
ordering dynamics of polymer blends within Edwards models (see Sec.\
\ref{sec:structural_models_particles}) \cite{mueller_smith_05,
daoulas_mueller_06}. The idea is to take snapshots of the density
configurations in regular intervals, and to let the chains move in
the fields created by these snapshots, {\em e.g.}, by kinetic Monte
Carlo simulations. If the fields were updated after every Monte Carlo
move, this would correspond to a regular simulation of the Edwards
model. Daoulas {\em et al} have shown that it is possible
to update much less often \cite{daoulas_mueller_06} without changing
the results. This makes the method very efficient and especially
suited for the use on parallel computers.

A similar idea had been put forward already in 2003 by Ganesan and
Pryamitsin \cite{ganesan_pryamitsyn_03}, in a less transparent
formulation that involves self-consistent fields, to study
stationary inhomogeneous polymer systems in an externally imposed
flow field. In the absence of flow, the model of Ganesan and
Pryamitsin is equivalent to that of M\"uller {\em et al.} A hybrid
method that combines kinetic Monte Carlo and the self-consistent
field formalism has also recently been proposed by B\"aurle and
Usami \cite{baeurle_usami_06}.

SCMF simulations are hampered by the general limitations of the
Edwards models -- chains can cross each other. However, there are
ways to introduce the dynamical effect of entanglements at least
in cases where the equilibrium configuration space is not affected
by topological constraints \cite{mueller_daoulas_08b}.

\subsubsection{Field-Based Dynamics}

\label{sec:dynamical_models_fields}

\paragraph{Dynamic Density Functional Theory (DDFT)}~\\
\nopagebreak
\vspace{-0.8\baselineskip}

Dynamic density functional theories (DDFT) are based on density
functionals for polymer systems, such as, {\em e.g.}, Eq.\
(\ref{eq:f_scf}). They supplement them by a model for their
dynamical evolution at nonequilibrium. For diffusive dynamics, the
dynamical equations in an imposed flow profile ${\bf v}$ have the
general form
\cite{fraaije_93,fraaije_vanvlimmeren_97,hasegawa_doi_97,kawakatsu_97}.
\begin{equation}
\label{eq:ddft}
 \frac{\partial \rho_i}{\partial t}
 + \nabla ({\bf v} \rho_i)
 = \int {\rm d}{\bf r}' \sum_{ij}
  \nabla_{\bf r} \Lambda_{ij}({\bf r},{\bf r}') \nabla_{{\bf r}'}
  \: \frac{\delta {\cal F}}{\delta \rho_j({\bf r}')}
 + \eta_{i}({\bf r},t),
\end{equation}
where $\Lambda_{ij}({\bf r},{\bf r}')$ is a generalized mobility, and
$\eta_{i}({\bf r},t)$ a Gaussian white noise with mean zero. If the
amplitude of the latter is very small or zero, one has 'mean-field
dynamics' and the system evolves towards a minimum of the free
energy functional (although not necessarily the global minimum). If
the noise is larger and satisfies the fluctuation dissipation
theorem,
\begin{equation}
 \langle \eta_i({\bf r},t) \eta_j({\bf r}',t') \rangle
 = - 2 \kBT \: \delta(t-t') \:
  \nabla_{\bf r} \Lambda_{ij}({\bf r},{\bf r}') \nabla_{{\bf r}'},
\end{equation}
the density configurations $\{\rho_i({\bf r})\}$ in an equilibrium
simulation (no flow, sufficiently long runs) will be distributed
according to $P[\rho] \propto \exp(- {\cal F}[\rho]/\kBT)$.

The kinetic Onsager coefficient $\Lambda_{ij}({\bf r},{\bf r}')$
depends on the microscopic dynamics in the system. Since it
characterizes the current of component $i$ in response to an
external force acting on component $j$, it is reasonable to assume
that it is proportional to the local density $\rho_i({\bf r})$. An
efficient choice which however disregards the connectivity of the
chains is thus $ \Lambda_{ij}({\bf r}, {\bf r}') = M_i \:
\rho_i({\bf r}) \: \delta({\bf r} - {\bf r}') \: \delta_{ij} $ for
compressible systems \cite{maurits_vanvlimmeren_97}, or
$\Lambda({\bf r}, {\bf r}') = M \: \rho_A  \rho_B \:
\delta({\bf r} - {\bf r}') $ for binary incompressible systems
(local dynamics) \cite{fraaije_vanvlimmeren_97}.

To account for the chain connectivity, one must include information
on the chain correlations. For example, for Rouse chains,
$\Lambda_{ij}({\bf r},{\bf r}')$ should be proportional to the pair
correlators $K_{ij}$ defined in Eq.\ (\ref{eq:pair_correlator})
\cite{kawakatsu_97,maurits_fraaije_97b}. At first sight, using such
complex Onsager coefficients in a simulation seems forbidding, but
thanks to a clever trick due to Maurits and Fraaije
\cite{maurits_fraaije_97b}, it becomes feasible
\cite{reister_mueller_01,mueller_schmid_05,he_schmid_06,he_schmid_06b,
he_schmid_08}.

The DDFT method has been extended, {\em e.g.}, to
account for viscoelastic \cite{maurits_zvelindovsky_98b} or
hydrodynamic effects \cite{maurits_zvelindovsky_98c,hall_lookman_06,
honda_kawakatsu_08}. A lattice version has also been proposed
\cite{mihajlovic_lo_05}.

\paragraph{Time Dependent Ginzburg-Landau (TDGL) Theories
and Cell Dynamics}~\\
\nopagebreak
\vspace{-0.8\baselineskip}

Like dynamic density functional theories, time-dependent
Ginzburg-Landau (TDGL) theories supplement a free energy functional
${\cal F}[\Phi]$ of an 'order parameter' field $\Phi({\bf r})$ by a
model for the dynamical evolution of $\Phi$. The TDGL theories of
interest in multiphase polymer systems mostly operate with locally
conserved order parameters and have the same general structure than
Eq.\ (\ref{eq:ddft}). For example, time-dependent Flory-Huggins-de
Gennes theories are used to study the demixing dynamics in polymer
blends, and time-dependent Ohta-Kawasaki theories to study the
ordering kinetics in copolymer systems. Discrete lattice versions of
TDGL theories are often referred to as 'cell dynamics' models.

We consider specifically the time-dependent Ohta-Kawasaki theory. 
Starting from the free energy functional (\ref{eq:ohta_kawasaki}) 
for melts of diblock AB copolymers and choosing an Onsager coefficient 
that describes local diffusive dynamics,
$\Lambda ({\bf r},{\bf r}') = M \: \delta({\bf r}-{\bf r}')$, we
obtain the dynamical equations \cite{oono_shiwa_87}
\begin{eqnarray}
\nonumber
 \frac{\partial \Phi_A}{\partial t}
 + \nabla ({\bf v} \Phi_A)
 &=& M \Delta \:
  \frac{\delta {\cal F}_{\mbox{\tiny FA}}}{\delta \Phi_A({\bf r})}
  + \eta({\bf r},t) \\
 &=& \frac{\rho_0}{N} \Big[ \Delta \Big(
    \frac{\partial {\cal W}}{\partial \Phi_A} - B \Delta \Phi_A \Big)
   - A (\Phi_A - \bar{\Phi}_A) \Big]
   + \eta({\bf r},t),
\label{eq:ok}
\end{eqnarray}
where $\bar{\Phi}_A$ is the total volume fraction of monomers A in
the system. Hence the somewhat awkward long-range 'Coulomb' term in
Eq.\ (\ref{eq:ohta_kawasaki}) becomes short range, and the dynamical
equations only depend on local terms. Because of the appealingly
simple structure of the final theory, Eq.\ (\ref{eq:ok}),
it is widely used for simulations of copolymer systems at equilibrium 
and under shear (see below). Bahiana and Oono have formulated a 
discrete version on a lattice \cite{bahiana_oono_90} which is equally 
popular in cell dynamics simulations.

TDGL simulations are much faster than molecular field simulations,
but of course, the underlying model is less accurate. Honda and
Kawakatsu have recently proposed a multiscale hybrid method that
combines the two approaches, using dynamic density functional input to 
improve on the accuracy of the TDGL model \cite{honda_kawakatsu_07}. 
Such hybrid approaches will presumably gain importance in the future.

\subsection{Applications}

\label{sec:applications}

After this overview over the main models used for simulations of
multiphase polymer systems, I will now illustrate them by reviewing
simulation work that has been done on homopolymer blends and
copolymer melts. I focus on simulations of generic models.
Atomistic studies or studies of bottom-up models are scarce and have
already been discussed earlier 
(Secs.\ \ref{sec:systematic_coarse_graining} and \ref{sec:atomistic}).

\subsubsection{Homopolymer Blends}

\label{sec:applications_homopolymers}

\paragraph{Bulk Properties}~\\
\nopagebreak
\vspace{-0.8\baselineskip}

I have already mentioned the pioneering simulations of binary
blends by Sariban and Binder and by Cifra {\em et al.}
\cite{sariban_binder_87,sariban_binder_88, cifra_karasz_88}.
Following up on this work, a number of studies, mainly by Binder and
coworkers, have considered the critical behavior of binary
blends \cite{deutsch_binder_92, deutsch_binder_93, binder_94,
mueller_binder_95, mueller_wilding_95, dadmun_waldow_99}. As
discussed in Sec.\ \ref{sec:fluctuations}, fluctuations shift the
critical demixing point and change the critical exponents from
mean-field-like to Ising-like. However, the region where the
critical behavior deviates from the mean-field prediction shrinks
with increasing chain length, and effective mean-field behavior
could be observed already for relatively moderate chain lengths
\cite{deutsch_binder_92}. Coming from the other end, field-based
Monte Carlo simulations of the Flory-Huggins-de Gennes functional
(\ref{eq:flory_huggins_degennes}) have confirmed that noise shifts
the coexistence curve and changes the critical behavior from
mean-field to Ising \cite{forrest_toral_94}.

Along with these studies of static blend properties, extensive work
has also been dedicated to the dynamics of demixing. On the
particle-based side, this was mainly investigated using kinetic
Monte Carlo \cite{sariban_binder_91, brown_chakrabarti_93,
reister_mueller_01}. Reister {\em et al} have compared results from
kinetic Monte Carlo simulations and different versions of the dynamic
density functional theory in a study of spinodal decomposition in
symmetric blends \cite{reister_mueller_01}. Other field-based
simulations have mainly relied on time-dependent Ginzburg Landau
models, which have the advantage that one can reach much later
stages of demixing. They were used to study demixing processes at
equilibrium \cite{takenaka_hashimoto_95, kielhorn_muthukumar_99,
cao_zhang_01, cheng_nauman_04, prusty_keestra_07} and under shear
\cite{he_naumann_97, zhang_zhang_01}. Particularly interesting
morphologies can be obtained if the dynamics in the two phases is
distinctly different, {\em i.e.}, one component becomes glassy
during the demixing process \cite{khalatur_khokhlov_94} or
crystallizes \cite{mehta_kyu_04, xu_chiu_06, zhou_shi_08}.

The particle-based studies mentioned so far have used coarse-grained
models of blends that demix explicitly for energetic reasons. A number 
of authors have explored other factors that are believed to affect the
chain miscibility with generic models \cite{mueller_99}, {\em e.g.},
the effect of nonrandom mixing \cite{kumar_94, yang_lu_94}, shape
disparity \cite{mueller_95}, stiffness disparity
\cite{weinhold_kumar_95}, different architectures
\cite{theodorakis_avgeropoulos_07}, and a different propensity
towards crystallization \cite{hu_mathot_03, ma_hu_07}.

\paragraph{Internal Interfaces}~\\
\nopagebreak
\vspace{-0.8\baselineskip}

In the miscibility gap, polymer blends have a highly inhomogeneous
structure with droplets of one phase dispersed in the other phase.
Their material properties are largely determined by the
properties of the interfaces separating the two phases. While the
distribution, size, and shape of the droplets depend on how the
blend has been processed, the interfaces separating them often have
time to reach local equilibrium and can be studied by means of
equilibrium simulations. The first study of an interface in a binary
blend was carried out by Reiter {\em et al.}
\cite{reiter_zifferer_90} in 1990. Since then, several
authors have investigated interfaces in symmetric
\cite{mueller_binder_95b, ypma_cifra_96, olaj_petrik_98,
lacasse_grest_98, werner_schmid_99, qian_lu_05} or asymmetric
\cite{mueller_werner_97, adhikari_auhl_02, adhikari_straube_03}
binary blends ({\em e.g.}, blends with stiffness and/or monomer size
disparities) by means of generic particle-based simulations.

The local interfacial structure is of interest because it determines
the mechanical stability of an interface. For example, the local
interfacial width gives the volume in which chains belonging to
different phases can entangle. On the other hand, we have already
discussed in Sec.\ \ref{sec:application_carelli} that interfaces
exhibit capillary waves, which are significant on all length scales.
This becomes apparent from the fact that the capillary-wave
contribution to the total width, as given by Eq.\ (\ref{eq:width_broadening}),
diverges both if the system size $L$ becomes very large and if the
microscopic coarse-graining length $a_0$ becomes very small.
Therefore the length scales of the capillary waves and those of the
local interfacial profiles cannot be separated clearly, and it is
not clear, {\em a priori}, whether the concept of a 'local
interfacial structure' is at all meaningful. This question has been
investigated by Werner {\em et al}\cite{werner_schmid_99}by
simulations of the bond-fluctuation model (see Sec.\
\ref{sec:structural_models_particles}). They demonstrated that it is
indeed possible to describe homopolymer interfaces consistently in
terms of a convolution of 'intrinsic' profiles with capillary wave
undulations. They also studied the influence of confinement both on
the capillary waves \cite{werner_schmid_97, werner_mueller_99} and
on the intrinsic interfacial width \cite{binder_mueller_99}.

Equilibrium simulations give information on the stability of
interfaces under mechanical stress, but in a rather indirect way. A
few authors have probed directly the rheological properties of
interfaces with nonequilibrium particle- and field-based simulation
methods, looking, {\em e.g.}, at shear thinning and interfacial slip
\cite{barsky_robbins_01,barsky_robbins_02, narayanan_pryamitsyn_04,
lo_mihajlovic_05}. Detailed simulations of nonequilibrium interfaces
are expensive, but with the development of new efficient simulation
algorithms and modern fast computers, they become feasible.

\paragraph{Surfaces and Films}~\\
\nopagebreak
\vspace{-0.8\baselineskip}

Another topic discussed intensely in the literature is the behavior
of blends in the vicinity of surfaces, and that of blends confined
to thin films. From a simulation point of view, these two situations
are identical, because surfaces are typically studied in slab
geometries ({\em i.e.}, with periodic boundaries in two directions
and free boundaries in the third).

A large amount of work has been dedicated to the somewhat artificial
situation of surfaces to which both types of monomers have exactly
equal affinity. Even though these surfaces are perfectly neutral,
one component will typically segregate to them: In incompatible
blends, the minority component aggregates at the surface in order
to reduce unfavorable contacts \cite{kumar_tang_94,
rouault_baschnagel_95, rouault_duenweg_96} (in contrast, the
minority chains are removed from the surface in miscible blends
\cite{cifra_karasz_92}). In blends of chains with different
stiffness, the stiffer chains are pushed towards the surface,
because they loose less entropy there \cite{yethiraj_kumar_94,
kumar_yethiraj_95}. For the same reason, linear chains aggregate at
surfaces in blends of linear and star polymers
\cite{batman_gujrati_08}.

Cavallo {\em et al} have systematically investigated the phase
behavior of films confined between neutral walls as a function of
the film thickness. If the film is made thinner, one observes a
crossover from three-dimensional to two-dimensional Ising behavior
\cite{cavallo_mueller_03, cavallo_mueller_05}. Fluctuation effects
in thin films are observed to be much stronger than in the bulk,
consistent with our discussion in Sec.\ \ref{sec:fluctuations}: In
two dimensions, the Ginzburg parameter no longer scales with the
chain length and stays finite for all chain lengths. A second
transition occurs when the film becomes so thin that {\em polymers}
are effectively two-dimensional, {\em i.e.}, they can no longer
pass each other. This reflects the theoretically expected
fundamental difference between the demixing behavior of overlapping
and non-overlapping two-dimensional polymers \cite{semenov_03}.

The more realistic situation of selective walls, to which one
component adsorbs preferentially, has been addressed as well by
different authors. In this case, the phase behavior is governed 
by wetting phenomena and capillary condensation 
\cite{pereira_wang_96, mueller_binder_98, mueller_binder_01}.

The studies discussed so far were based on particle simulations. A
few authors have used field-based simulations to explore dynamical
aspects of phase separation in thin polymer blends. Morita {\em et
al} have studied the interplay of spinodal decomposition and
interfacial roughening due to droplet formation with dynamic density
functional simulations \cite{morita_kawakatsu_01}. 
Shang {\em et al} have used a time-dependent Ginzburg-Landau approach 
to study the spinodal phase separation of a thin film on a 
heterogeneous substrate \cite{shang_kazmer_08}.

\subsubsection{Copolymer Systems}

\label{sec:applications_copolymers}

\paragraph{Copolymers as Compatibilizers}~\\
\nopagebreak
\vspace{-0.8\baselineskip}

Copolymers were originally designed as natural surfactant molecules
that increase the miscibility of incompatible homopolymers and
enhance their interfacial properties. They are usually much more
expensive than their respective homopolymers, but adding a small
amount of copolymer can already improve the properties of the
homopolymer blend significantly.

A number of researchers have considered the effect of copolymers on
the demixing transition for different copolymer architectures
\cite{dadmun_96, dadmun_waldow_99, dai_song_04}. Dadmun and Waldow
\cite{dadmun_waldow_99} have pointed out that copolymers not only
shift the transition point towards higher values of the
Flory-Huggins parameter $\chi$, but also change the critical
exponents {\em} via a Fisher-renormalization mechanism. In the
phase-segregated regime, copolymers aggregate to the interface,
reduce the interfacial tension, and enlarge the interfacial width.
The interfacial structure of homopolymer interfaces with adsorbed
copolymers has been explored in detail by several authors
\cite{dadmun_96, werner_schmid_96, werner_schmid_99b,
liang_99,kim_jo_99, qian_lu_05}.

Milner and Xi have noted in 1996 that the main compatibilizing effect 
of copolymers probably has a kinetic origin \cite{milner_xi_96}: 
Copolymers reduce the rate of droplet coalescence during the processing 
of the blend {\em via} a Marangoni effect: If the copolymer 
concentration drops somewhere at the surface of a droplet,
the surface tension increases locally. This induces flow in the 
direction of the weak point. Hence the copolymer film stabilizes 
itself kinetically, much like a soap film. In addition, the copolymer 
blocks stretching into the bulk form repulsive coronae. Experimental 
studies suggest that the resulting steric repulsion between droplets 
may be even more prohibitive for droplet coalescence than the Marangoni 
effect \cite{lyu_jones_02}. Kim and Jo have studied the
influence of copolymers on the dynamics of demixing in a series of
works \cite{jo_kim_96, kim_jo_96, kim_jo_97, kim_jo_98,ko_kim_00},
but they used kinetic Monte Carlo, hence they did not include
hydrodynamic effects and could not study the effect of the Marangoni
flow.

If one increases the copolymer concentration beyond a certain
threshold, the macro\-phase separated phase at high $\chi N$ eventually
gives way to a microphase separated phase (see, {\em e.g.}, the
phase diagram shown in Fig.\ \ref{fig:me_phases}). A few authors
have explored the full phase behavior of ternary systems
\cite{mueller_schick_96, poncela_rubio_03, duechs_ganesan_03,
duechs_schmid_04}. In ternary systems containing A and C
homopolymers and ABC triblock copolymers, a whole zoo of new
tricontinuous gyroid phases can be observed \cite{dotera_02}.

\paragraph{Pure Bulk Copolymer Melts}~\\
\nopagebreak
\vspace{-0.8\baselineskip}

The propensity of copolymers to self-assemble into complex
mesostructures makes them attractive for various micro-
and nanotechnological applications, which is why pure copolymer
melts have become interesting in their own right. Many simulation
studies are now concerned with the properties of pure copolymer
melts.

Early studies of fluctuations and chain correlations near the
order/disorder transition (ODT), coming from the disordered side
\cite{fried_binder_91, fried_binder_91b, binder_fried_93,
pakula_karatasos_97} have reproduced the ODT-singularity in the
structure factor and revealed the dumbbell structure of the chains
mentioned earlier \cite{fried_binder_91}. Later, intense work has
been devoted to studying ordered lamellar structures below the ODT
in melts of symmetrical diblock copolymers, and analyzing them with
respect to their structure, their dynamics, and their fluctuations
\cite{weyersberg_vilgis_93, larson_94, larson_95, pan_shaffer_96,
hoffmann_sommer_97, banaszak_clarke_99, jo_jang_99, murat_grest_98,
murat_grest_99, wang_nealey_02, yang_winnik_05, srinivas_swope_07}.
Simulation studies of asymmetric diblock copolymer melts have also
reproduced most of the other mesophases -- the cylindrical phase,
the bcc sphere phase, and even the gyroid phase
\cite{hoffmann_sommer_97b, groot_madden_98, groot_madden_99,
besold_hassager_99, horsch_zhang_04, martinez_escobedo_05,
martinez_escobedo_06}.

Locating the actual position of the ODT accurately is difficult,
especially in lattice models \cite{micka_binder_95}, since the
natural periodicity of the structures is in general incompatible
with the box size. This results in complex finite size artefacts.
Nevertheless, phase diagrams have been obtained in recent years,
\cite{schultz_hall_02, matsen_griffiths_06, beardsley_matsen_08}. 
Particle-based and field-based simulations have provided evidence 
that for symmetrical diblock copolymers ($f=1/2$), the transition 
to the lamellar phase is shifted \cite{ganesan_fredrickson_01, 
vassiliev_matsen_03}, compared to the mean-field phase diagram, 
and becomes first order \cite{vassiliev_matsen_03, mueller_daoulas_08, 
lennon_katsov_08}, in agreement with the theoretical expectation
\cite{fredrickson_helfand_87} (see Sec.\ \ref{sec:fluctuations}).

\begin{figure}
\centerline{
 \includegraphics[width=0.48\textwidth]{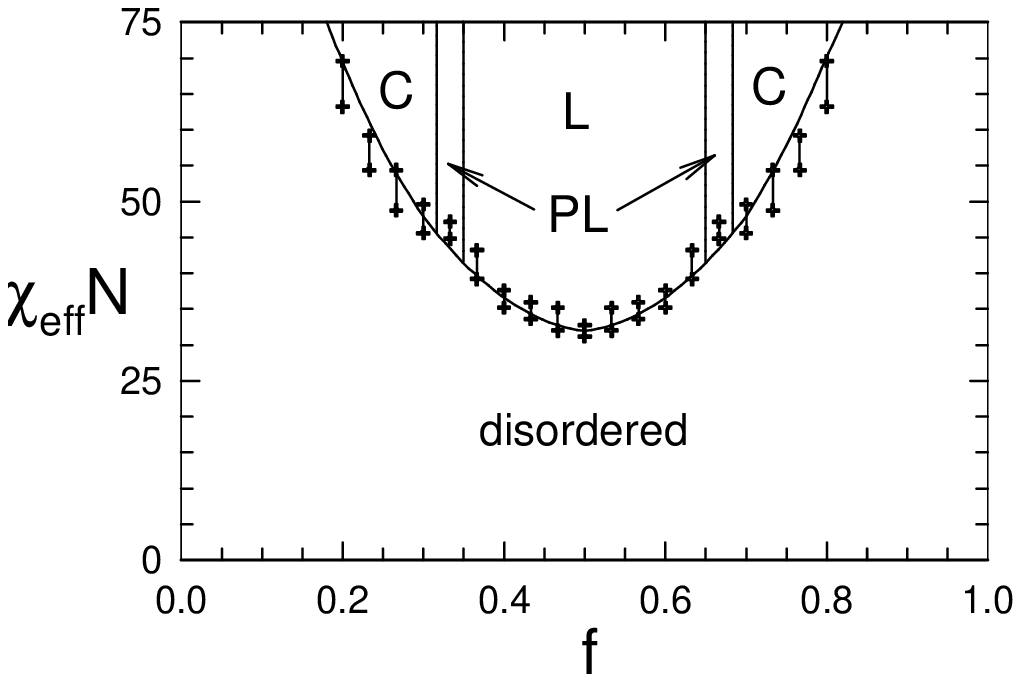}
 \includegraphics[width=0.48\textwidth]{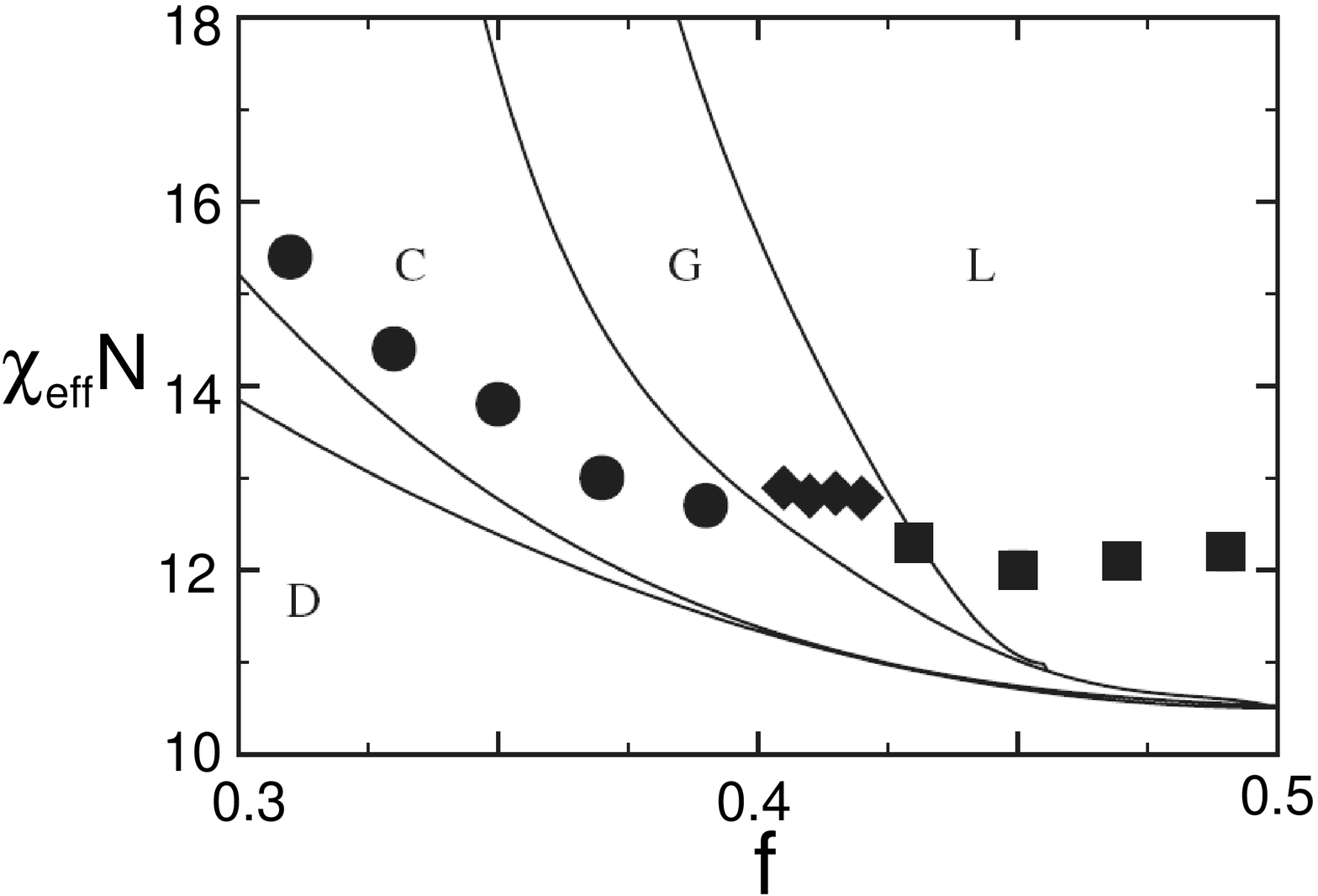}
 }
\caption{\label{fig:diblock_2} Phase diagrams for diblock copolymer
melts as obtained from Monte Carlo simulations of a lattice model
(chain length $N=30$) by Matsen {\em et al.} (left) 
\protect\cite{matsen_griffiths_06} and from 
field-theoretical Complex-Langevin simulations (Ginzburg parameter 
$C=50$) by Lennon {\em et al} (right) \protect\cite{lennon_katsov_08}. 
Symbols L,C,G correspond to diblock phases of Fig.\ 
\protect\ref{fig:structures_cops}; in addition, the Monte Carlo 
phase diagram features a perforated lamellar (PL) phase.  
Left figure: Courtesy of Mark W. Matsen, adapted from Matsen,
Griffiths, Wickham, and Vassiliev, J. Chem. Phys. {\bf 124}, 
024904 (2006).
Right figure: Reprinted with permission from Lennon, Katsov,
, and Fredrickson, Phys. Rev. Lett. {\bf 101}, 138302, 2008.
Copyright (2008) by the American Physical Society.
}
\end{figure}

Two examples of phase diagrams obtained with different simulation
methods are shown in Fig.\ \ref{fig:diblock_2}: One was calculated
by Matsen {\em et al} \cite{matsen_griffiths_06} using Monte Carlo
simulations of copolymers with length $N=30$ in a simple lattice
model (left), and one by Lennon {\em et al.}\cite{lennon_katsov_08}
using field-theoretical calculations (right) at Ginzburg parameter
$C=50$. Both phase diagrams significantly improve on the SCF phase
diagram of Fig.\ \ref{fig:diblock_1} (right) in the weak segregation
regime, and reproduce the main qualitative features of the
experimental phase diagram (Fig.\ \ref{fig:diblock_1}, left): The
transitions are first order everywhere. The ODT is shifted to higher
$\chi N$. Direct phase transition between the disordered phase and
the complex mesophase (PL or G, respectively) or the lamellar phase
are possible for a range of copolymer block fractions $f$. The phase
diagrams even reproduce a small hump of the ODT transition line at
the boundary to the complex mesophase (PL or G), which is also
observed experimentally. The only 'problem' with the Monte Carlo phase 
diagram is that it features a perforated lamellar (PL) phase instead 
of a gyroid (G) phase. This may be a finite size artefact, as suggested 
by the authors, or a property of the lattice model under consideration. 
(Gyroid phases have been found in lattice models
\cite{martinez_escobedo_05}, but it should be noted that the free 
energies of the PL and the G phase are very close according to SCF 
calculations.) Nevertheless, the simulations demonstrate convincingly 
that the discrepancies between the experimental phase diagram and the 
SCF phase diagram shown in Fig.\ \ref{fig:diblock_1} can to a large 
extent be attributed to the effect of fluctuations.

In recent years, researchers have also begun to simulate melts of
more complex copolymers, {\em e.g.}, starblock copolymers
\cite{gemma_hatano_02, xu_feng_06, xu_feng_08,
he_huang_02,he_huang_03}, rod-coil copolymers,
\cite{alsunaidi_denotter_04, pryamitsyn_ganesan_04} diblock
copolymers with one crystallizing component \cite{hu_frenkel_05,
hu_05, hu_karssenberg_06, wang_hu_06, cho_wang_06, qian_cai_08},
triblock copolymers \cite{ko_kim_01, banaszak_woloszczuk_02,
huh_jo_02, song_shi_08}, or random copolymers
\cite{houdayer_mueller_02, houdayer_mueller_04}. An interesting
study on diblock copolymers with one {\em amphiphilic} block has
recently been presented by Khokholov and Khalatur
\cite{khokhlov_khalatur_08}. Since the amphiphilic block favors
interfaces, the morphology of the mesophases changes completely and
is characterized by thin channels and slits. Lay {\em et al.} and
Palmer {\em et al} have studied the computationally challenging
problem of microphase separation in randomly crosslinked binary
blends \cite{lay_sommer_00, lay_sommer_00b, palmer_lastoskie_04},
and also the inverse problem, to which extent ordered copolymer
structures can be stabilized by cross-linking \cite{lay_sommer_99}.

A large amount of simulation work on copolymer melts has been done 
with time-dependent Ginzburg-Landau approaches \cite{oono_bahiana_88,
chakrabarti_toral_91, shiwa_taneike_96, komura_fukuda_96,
qi_wang_96, qi_wang_97, zhang_zhang_97, hamley_00,
hamley_stanford_00, ren_hamley_01, vega_harrison_05, ohta_ito_95,
ohta_motoyama_96, ito_98, zhang_jin_06}. These studies have mostly
addressed dynamical questions, {\em i.e.}, the kinetics of ordering,
disordering processes in pure melts \cite{shiwa_taneike_96,
komura_fukuda_96, qi_wang_96, qi_wang_97, zhang_zhang_97, hamley_00,
hamley_stanford_00, ren_hamley_01, vega_harrison_05} and in mixtures
containing copolymers \cite{ohta_ito_95,ohta_motoyama_96, ito_98,
zhang_jin_06}. Already in pure diblock copolymer melts,
ordering/disordering processes were found to proceed {\em via}
intricate pathways that involve nontrivial intermediate states ({\em
e.g.}, the perforated lamellar state which is only metastable at
equilibrium). In mixtures, the interplay of macrophase and
microphase separation leads to a wealth of new transient
morphologies \cite{ohta_ito_95,ohta_motoyama_96, ito_98,
zhang_jin_06} (see also Ref.\ \citeonline{morita_kawakatsu_02}).

\paragraph{Confined Copolymer Melts}~\\
\nopagebreak
\vspace{-0.8\baselineskip}

In recent years, there has been growing interest in confined
copolymer systems. The first studies have explored the ordering of
copolymers melts in thin films between neutral walls
\cite{sommer_hoffmann_99, feng_ruckenstein_02, song_li_07,
alexanderkatz_fredrickson_07} or general (selective) walls
\cite{kikuchi_binder_93, kikuchi_binder_94, geisinger_mueller_99,
geisinger_mueller_99b, huinink_brokkenzijp_00, wang_yan_00,
feng_ruckenstein_02b, pereira_02, feng_liu_04, knoll_lyakhova_04,
lyakhova_sevink_04}, both with particle-based models and dynamic
density field theories. Triblock copolymers \cite{szamel_mueller_03,
ludwigs_boker_03, horvat_lyakhova_04, lyakhova_horvat_06,
huang_liu_06, xiao_huang_07} and copolymer blends \cite{song_li_07}
have also been considered. The dynamics of copolymer ordering in
confinement was studied with time-dependent Ginzburg-Landau methods
\cite{brown_chakrabarti_94, brown_chakrabarti_95, feng_liu_04}. A
series of papers have dealt with the technologically relevant
question, whether and how surface patterns can be transferred into
copolymer films, \cite{pereira_williams_00, wang_nath_00,
wang_yan_00, jayaraman_hall_05,  daoulas_mueller_06c,
daoulas_mueller_08, chen_peng_07}.

The current focus shifts to nanocylindrically or
spherically confined blends \cite{xu_wu_06, feng_ruckenstein_06,
feng_ruckenstein_06b, feng_liu_06, feng_ruckenstein_07, feng_liu_07,
feng_liu_08, feng_ruckenstein_08, wang_07, xiao_huang_07b,
han_cui_08}. A variety of new structures can already be observed in
systems of diblock copolymers confined to nanocylinders, {\em e.g.},
mesh structures, single and double helices
\cite{feng_ruckenstein_06}. For triblock copolymers, the spectrum is
even more diverse \cite{feng_ruckenstein_07}. The confined
structures depend on the bulk structure and on the shape of the
confining channels. Li and coworkers have shown that possible
morphologies can be screened efficiently with the method of
simulated annealing \cite{yin_sun_04, yin_sun_06, yu_sun_06,
yu_sun_07, yu_li_07, chen_liang_08}.

\paragraph{Copolymers under Shear}~\\
\nopagebreak
\vspace{-0.8\baselineskip}

Finally in this section, I briefly mention the large amount of work
that has addressed the effect of shear on the microstructure of
copolymer systems, with both field-based and particle-based methods
\cite{ohta_enomoto_93,kodama_doi_96, zvelindovsky_vanvlimmeren_98,
zvelindovsky_sevink_98, zhang_manke_00, hamley_01, ren_hamley_02,
sevink_zvelindovsky_04, feng_ruckenstein_04, honda_kawakatsu_06,
rychkov_05, liu_zhong_05, narayanan_pryamitsyn_06, li_tang_08,
you_chen_07, pinna_zvelindovsky_06, pinna_zvelindovsky_08}. Special
attention was given to the phenomenon that lamellae reorient under
shear \cite{ren_hamley_02, fraser_denniston_05, fraser_denniston_06,
liu_qian_06, lisal_brennan_07}. Experimentally, it is observed that
lamellae orient parallel to the shear flow at low shear rates, and
perpendicular to the shear flow at high shear rates. Simulations
give conflicting results. The parallel state consistently becomes
unstable at high shear rates, but its relative stability at low
shear rates (or results on the indicators of relative stability such
as the entropy production) seems to depend on the type of model
\cite{fraser_denniston_06,lisal_brennan_07}.

\section{Future Challenges}

The equilibrium theory of fluid polymer mixtures is fairly advanced.  
Thanks to the universal properties of polymers, it requires
relatively little input information ($\chi$ parameter etc.) to be
predictive at a quantitative level. However, if one goes beyond
equilibrium and beyond the fluid state, the situation is much less
satisfying. For example, very little work has been done on
crosslinked polymer blends \cite{lay_sommer_99, lay_sommer_00,
lay_sommer_00b, palmer_lastoskie_04}, even though they are common in
applications. Physical crosslinks can be established if domains
crystallize or become glassy. As we have seen above, research on
blends with one crystallizing or glassy component is also rather
scarce.

Another important issue is the influence of the blend processing 
on the properties of the resulting materials, {\em i.e.}, the structure 
of phase separating blends under shear. The mechanical properties of 
immiscible blends depend crucially on their microscopic morphologies, 
{\em i.e.}, the sizes and shapes of droplets, which in turn depend 
on the manufacturing process. Theoretical nonequilibrium state diagrams 
that relate the processing conditions (shear rates, geometry, 
copolymer concentration etc.) with final morphologies are still missing. 
Since the relevant length scales are relatively large and hydrodynamics 
are important, the simulation method of choice should be a cell 
dynamics method that incorporates hydrodynamics, {\em e.g.}, a 
Lattice-Boltzmann method \cite{succi_book}. Methods that combine 
Ginzburg-Landau functionals for immiscible fluids with Lattice-Boltzmann 
models for Newtonian fluids have been developed \cite{orlandini_swift_95} 
and used to study demixing processes at rest \cite{wagner_yeomans_98} 
and under shear \cite{wagner_yeomans_97, wagner_yeomans_99,
martys_douglas_01}. Giraud and coworkers have proposed a
Lattice-Boltzmann method for viscoelastic fluids, which is more
suitable to describe polymers \cite{giraud_dhumieres_98}, and
carried out first simulations of viscoelastic liquid mixtures
\cite{wagner_giraud_00}. Nevertheless, the whole field is still in
its infancy.

As the field of polymer simulations reaches maturity, the bottom-up
modeling approach (Sec.\ \ref{sec:systematic_coarse_graining}) will
gain importance. So far, the vast majority of theoretical
and simulation studies of (co)polymer blends was based on generic
model systems. One or two decades from now, realistic simulations of
specific polymer blends will probably be equally, if not more
common. One of the major challenges in this context is to develop hybrid
multiscale methods that combine different levels of coarse-graining,
{\em i.e.}, use a relatively coarse basic model and fine-grain
selectively in interesting regions of the materials ({\em e.g.},
interfaces) \cite{prapotnik_dellesite_08}.

\subsection*{Acknowledgments}

The author thanks Mark W. Matsen for introducing her to the
self-consistent field theory a long time ago and for providing
Figs.\ \ref{fig:diblock_1} and \ref{fig:diblock_2} (left), and Glenn
H. Fredrickson for discussions and for the permission to show Fig.\
\ref{fig:diblock_2} (right). She has benefitted from collaborations
and/or discussions with J\"org Baschnagel, Kurt Binder, Dominik
D\"uchs, Burkhard D\"unweg, Avi Halperin, Venkat Ganesan, Kurt
Kremer, Marcus M\"uller, Wolfgang Paul, Ulf Schiller, Jens Smiatek,
Jens-Uwe Sommer, Andreas Werner, and many others. Financial support
from the German Science Foundation (DFG) is gratefully acknowledged.

\begin{small}
\parsep=0cm
\itemsep=0cm
\bibliographystyle{wileyj}
\bibliography{handbook}
\end{small}

\end{document}